%% file: main.tex
\documentclass[preprint]{elsarticle}

\usepackage{ltablex}
\usepackage{graphicx}
\usepackage{url}
\usepackage{doi} 
\usepackage{tikz}
\usepackage{pgfplots}
\usetikzlibrary{shapes,arrows}
\usepackage{bchart}
\usepackage{rotating}
\usepackage{booktabs}
\usepackage{array}
\usepackage{caption}
\usepackage[nomessages]{fp}
\usepackage{subcaption}
\usepackage{textcomp} 

\usepackage{multicol}
\usepackage{multibib} 
\newcites{R}{Selected Papers}
\usepackage{hyperref}

\usepackage{multibib}
\newcites{r}{Selected Publications}

\journal{Information and Software Technology}

\pgfplotsset{compat=1.15}

\begin{document}
\begin{frontmatter}

\title{Controlled Experimentation in Continuous Experimentation: Knowledge and Challenges}

\author[1]{Florian Auer\corref{cor1}}
\ead{florian.auer@uibk.ac.at}
\cortext[cor1]{Both authors contributed equally to this work and are corresponding authors. }

\author[2]{Rasmus Ros\corref{cor1}}
\ead{rasmus.ros@cs.lth.se}

\author[1]{Lukas Kaltenbrunner}
\ead{lukas.kaltenbrunner@uibk.ac.at}

\author[2]{Per Runeson}
\ead{per.runeson@cs.lth.se}

\author[1,3]{Michael Felderer}
\ead{michael.felderer@uibk.ac.at}

\address[1]{University of Innsbruck, Austria}
\address[2]{Lund University, Sweden}
\address[3]{Blekinge Institute of Technology, Sweden}

\begin{abstract}

\emph{Context:}
Continuous experimentation and A/B testing is an established industry practice that has been researched for more than 10 years. Our aim is to synthesize the conducted research. 

\emph{Objective:}
We wanted to find the core constituents of a framework for continuous experimentation and the solutions that are applied within the field. Finally, we were interested in the challenges and benefits reported of continuous experimentation. 

\emph{Method:}
We applied forward snowballing on a known set of papers and identified a total of 128 relevant papers. Based on this set of papers we performed two qualitative narrative syntheses and a thematic synthesis to answer the research questions. 

\emph{Results:}
The framework constituents for continuous experimentation include experimentation processes as well as supportive technical and organizational infrastructure. 
The solutions found in the literature were synthesized to nine themes, e.g. experiment design, automated experiments, or metric specification. 
Concerning the challenges of continuous experimentation, the analysis identified cultural, organizational, business, technical, statistical, ethical, and domain-specific challenges. 
Further, the study concludes that the benefits of experimentation are mostly implicit in the studies. 

\emph{Conclusions:}
The research on continuous experimentation has yielded a large body of knowledge on experimentation. The synthesis of published research presented within include recommended infrastructure and experimentation process models, guidelines to mitigate the identified challenges, and what problems the various published solutions solve.

\end{abstract}


\end{frontmatter}
\section{Introduction}

Deciding which feature to build is a difficult problem for software development organizations. The effect of an idea and its return-on-investment might not be clear before its launch. Moreover, the evaluation of an idea might be expensive. Thus, decisions are based on experience or the opinion of the highest paid person~\cite{kohavi2007practical}. 
Similarly difficult is the assessment of technical changes on products. It can be difficult to predict the effect of a change on software quality, as evidenced by the extensive research on e.g. defect prediction~\cite{fenton1999critique,wahono2015systematic} or software reliability estimation~\cite{ronchieri2018metrics}. Moreover, there are cases in which it is not feasible to test for all necessary scenarios, e.g. in all relevant software and hardware combinations.

Continuous experimentation (CE) addresses these problems.
It provides a method to derive information about the effect of a change by comparing different variants of the product to the unmodified product (i.e. A/B testing). This is done by exposing different users to different product variants and collecting data about their behavior on the individual variants. Thereafter, the gathered information allows making data-driven decisions and thereby reducing the amount of guesswork in the decision making.

In 2007, Kohavi et al. \citer{kohavi2007practical} published an experience report on experimentation at Microsoft and provided guidelines on how to conduct so-called \emph{controlled experiments}. It is the seminal paper about continuous experimentation and thus represents the start of the academic discussion on the topic. Three years later, a talk from the Etsy engineer Dan McKinley~\cite{mckinley2012design} gained momentum in the discussion. In the talk, the term \emph{continuous experimentation} was used to describe their experimentation practices. Other large organizations, like Facebook \citer{feitelson2013development} and Netflix \citer{uribe2015netflix}, which adopted data-driven decision making~\citer{kohavi2013online}, shared their experiences~\citer{borodovsky2011ab} and lessons learned~\citer{kohavi2009online} about experimentation over the years with the research community. In addition, researchers from industry as well as academia developed methods, models and optimizations of techniques that advanced the knowledge on experimentation. 

After more than ten years of research, numerous work has been published in the field of continuous experimentation, including work on problems like the definition of an experimentation process~\citer{fagerholm2017right}, how to build infrastructure for large-scale experimentation~\citer{gupta2018anatomy}, how to select or develop metrics \citer{machmouchi2016principles}, or the considerations necessary for various specific application domains~\citer{eklund2012architecture}. 

The purpose of this systematic literature review is threefold. First, to synthesize the models suggested by the research community to find characteristics of an essential framework for experimentation. This framework can be used by practitioners to identify elements in their experimentation framework. Second, to synthesize the various technical solutions that have been applied. In this inquiry, we also include to what degree the solutions are validated.
Finally, to summarize and categorize the challenges and benefits with continuous experimentation. Based on this the following four research questions are addressed in this work:

\begin{itemize}
    \item[]\textbf{RQ1:} What are the core constituents of a CE framework?
    \item[]\textbf{RQ2:} What technical solutions are applied in what phase within CE?
    \item[]\textbf{RQ3:} What are the challenges with CE?
    \item[]\textbf{RQ4:} What are the benefits with CE?
\end{itemize}

The research method of this study is based on two independently conducted mapping studies~\citer{auer2018current, ros2018continuous2}. We extended and validated the studies by cross-examining the included studies. Thereafter, we applied two qualitative narrative syntheses and a thematic synthesis on the resulting set of papers.

In the following Section \ref{sec:background} an overview of continuous experimentation and related software practices is given. Next, Section \ref{sec:research_method} describes the research method applied and Section \ref{sec:results} presents the results of the research. In Section \ref{sec:discussion} the findings are discussed. Finally, Section \ref{sec:conclusions} summarizes the research.

\section{Background}
\label{sec:background}

In this section we present an overview of continuous experimentation and related continuous software engineering practices. Further, we summarize our two previously published mapping studies.
For the novice reader, we recommend Fagerholm et al.'s descriptive model of continuous experimentation~\citer{fagerholm2017right}, or Kohavi et al.'s tutorial on controlled experiments~\citer{kohavi2008controlled}, which is a more hands on introduction for continuous experimentation. 

\subsection{Continuous software engineering}
\label{sec:background_continuous_software_engineering}
In their seminal paper on controlled experiments on the web from 2007, Kohavi et al.~\citer{kohavi2007practical} explain how the ability to continuously release new software to users is crucial for efficient and continuous experimentation, which is now known as continuous delivery and continuous deployment. Together with continuous integration, these are the three software engineering practices that allow software companies to release software to users rapidly and reliably~\cite{shahin2017continuous} and are fundamental requirements for continuous experimentation.

\emph{Continuous integration} entails automatically merging and integrating software from multiple developers. This includes testing and building an artifact, often multiple times per day.
\emph{Continuous delivery} is the process by which software is ensured to be always in state to be ready to be deployed to production. Successful implementation of continuous integration and delivery should join the incentives of development and operations teams, such that developers can release often and operations get access to powerful tools. This has introduced the DevOps~\cite{ebert2016devops} role in software engineering with responsibility for numerous activities: testing, delivery, maintenance, etc. 
Finally, with \emph{continuous deployment}, the software changes that successfully make it through continuous integration and continuous delivery can be deployed automatically or with minimal human intervention. Continuous deployment facilitates collection of user feedback through faster release cycles~\cite{fabijan2015customer,yaman2016customer}. With faster release cycles comes the ability to release smaller changes, the smaller the changes are the easier it becomes to trace feedback to specific changes.

Fitzgerald and Stol~\citer{fitzgerald2017continuous} describe many more continuous practices that encompass not only development and operations, but also business strategy; among them continuous innovation and continuous experimentation. Experiments are means to tie development, quality assurance, and business together, because experiments provide a causal link between software development, software testing, and actual business value. Holmstr{\"o}m Olsson et al.~\cite{olsson2012climbing} describe how ``R\&D as an experiment system'' is the final step in a process that moves through the continuous practices. 

\subsection{Continuous experimentation}
\label{sec:background_process}
The process of conducting experiments in a cycle is called \emph{continuous experimentation}. The reasoning is that the results of an experiment often begets further inquires. Whether the original hypothesis was right or wrong, the experimenter learns something either way. This learning can lead to a new hypothesis which is subject to a new experiment. This idea of iterative improvement is known since long from the engineering cycle or from iterative process improvements, as explained in the models Plan-Do-Check-Act~\cite{deming1986out} or quality process improvement paradigm (QIP)~\cite{basili1985quantitative}. The term ``continuous experimentation'' as used by software engineering researchers refers to a holistic approach~\citer{fitzgerald2017continuous} which spans a whole organization. It considers the whole software life-cycle, from business strategy and planning over development to operations.

Some authors have included many methods of gathering feedback in continuous experimentation~\citer{fagerholm2017right,lindgren2016raising}, including qualitative methods and data mining. These methods are not the focus of this work, though they are also valuable forms of feedback~\cite{bosch2015user,yaman2016customer}. For example, qualitative focus groups in person with selected users can be used early in development on sketches or early prototypes. The human-computer interaction research field has studied this extensively---recently under the name of \emph{user experience research}---and it has also been the subject of software engineering literature reviews in combination with agile development~\cite{jurca2014integrating,salah2014systematic}. In contrast to the qualitative methods, a controlled experiment requires a completed feature before it can be conducted. It is focused on quantitative data, thus cannot easily answer questions on the rationale behind the results, as qualitative methods can. As such, these methods compliment each other, but they are different in terms of methodology, infrastructure, and process. We discuss the qualitative methods through the lens of controlled experimentation in Section~\ref{sec:results_qualitative}. 

A \emph{randomized controlled experiment} (or A/B test, bucket test, or split test) is a test of an idea or a \emph{hypothesis} in which variables are systematically changed to isolate the effects. Because the outcome of an experiment is non-deterministic, the experiment is repeated with multiple \emph{subjects}. Each subject is randomly assigned to some of the variable settings. The goal of the experiment is to investigate whether changes in the variables have a causal effect on some output value, usually in order to optimize it. In statistical terminology, the variable that is manipulated is called the independent variable and the output value is called the dependent variable. The effect that changing the independent variables has on the dependent variable can be expressed with a statistical \emph{hypothesis test}. A significance test involves calculating a p-value\footnote{T-test is often used to compare whether the mean of two groups are equal, based on the t-score $t=\frac{\bar{x}_1-\bar{x}_2}{s/\sqrt{n}}$, where $\bar{x}$ is mean, $s$ is the standard deviation, and $n$ is the number of data points. The p-value is derived from the t-score through the t-distribution.} and the hypothesis is validated if the p-value is below a given \emph{confidence level}, often $95\%$. In addition, properly conducting a controlled experiment requires a power calculation\footnote{A simple approximate power calculation~\cite{kohavi2007practical} for fixed $95\%$ confidence level and $90\%$ statistical power is $n = (4rs/\Delta)^2$, where $n$ is the number of users, $r$ is the number of groups, $s$ is the standard deviation, and $\Delta$ is the effect to detect.} to decide experiment duration.

In software engineering, a controlled experiment is often used to validate a new product feature, in that case the independent variable is whether a previous baseline feature or the new feature should be used. These are sometimes called control and test group, or the A and B group, in which case the \emph{experiment design} is called an A/B test. In an A/B test, only one variable is changed; other experiment designs are possible~\citer{fisher1937design,kohavi2008controlled} but rarely used~\cite{ros2018continuous2}. To optimize software configuration settings is another use of controlled experiments in software engineering~\citer{letham2019constrained}. The dependent variable of the experiment is some measurable metric, designed with input from some business or customer needs. If there are multiple metrics involved with the experiment, then an overall evaluation criteria (OEC)~\cite{roy2001design} can be used, which is the most important metric for deciding on the outcome of the experiment. The subjects of the experiments are usually users, that is, each user provides one or more data points. In some cases the subjects are hardware or software parameters, for example, when testing optimal compiler settings.

\begin{figure}
\centering
\small
\input{figure-ce-process.tex}
\caption{Continuous experimentation process overview in five phases. A hypothesis is prioritized and implemented as a minimum viable product, then an experiment is designed and conducted that evaluates the software change, finally a decision is made to continue or pivot to another feature. This simplified process is based on the RIGHT model. The roles involved in each phase are shown to the left.}
\label{fig:ce_process}
\end{figure}
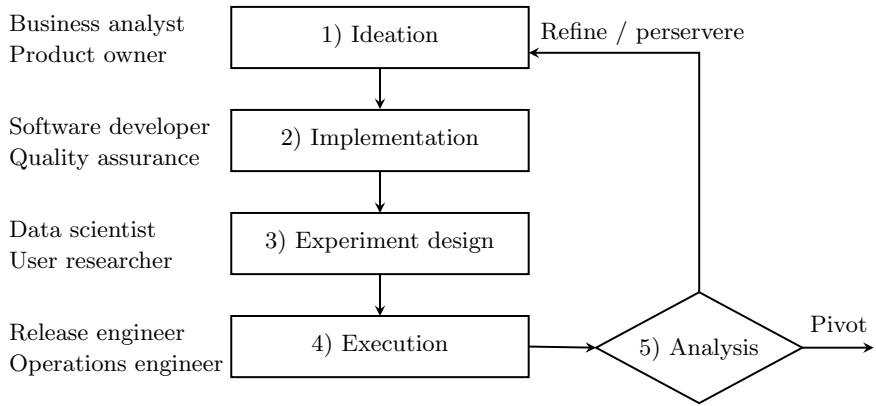

The process of continuous experimentation (see Fig.~\ref{fig:ce_process}) has similarities to the tradition from science in software engineering research~\cite{wohlin2012experimentation} and elsewhere~\cite{fisher1937design}. However, we base the following process on the RIGHT model by Fagerholm et al.~\citer{fagerholm2017right}. There are five main phases of the process. 1) In the \emph{ideation} phase hypotheses are elicited and prioritized. 2) \emph{Implementation} of a minimum viable product or feature (MVP) that fulfill the hypothesis follows. 3) Then, a suitable \emph{experiment design} with an OEC is selected. 4) \emph{Execution} involves release engineers deploying the product into production and operations engineers monitoring the experiment in case something goes wrong. Finally, 5) an \emph{analysis} is conducted with either statistical methods by data scientists or by qualitative methods by user researchers. If the results are satisfactory the feature is included in the product and a new hypothesis is selected so the product can be further refined. Otherwise, a decision must be made if to persevere and continue the process or if a pivot should be made to some other feature or hypothesis. Lastly, the results should be generalized into knowledge so the experience gained can be used to inform future hypotheses and development on other features.  

Many of the papers included in this study are on improved analysis methods. One such direction that need additional explanation is \emph{segmentation}. It is used in marketing to create differentiated products for different segments of the market. In the context of experiments it is used to calculate metrics for various slices of the data in, e.g. gender, age groups, or country. Experimentation tools usually perform automated segmentation~\citer{gupta2018anatomy} and can, for example, send out alerts if a change affects a particular user group adversely.

\subsection{Previous Work}
\label{sec:previous_work}

Prior to this literature review, two independent mapping studies~\citer{ros2018continuous,auer2018current} were conducted by the authors. Although both studies were in the context of continuous experimentation, their objectives differed.

In their mapping study \citer{ros2018continuous}, Ros and Runeson provided a short thematic synthesis of the topics in the published research and examined the context of the research in terms of reported organisations and types of experiments that were conducted. They found that there is a diverse spread of organisations of company size, sector, etc. Although, continuous experimentation for software that does not require installation (e.g. websites) was more frequently reported. Concerning the experimentation treatment types, the authors found more reports about visual changes than algorithmic changes. In addition, the least common type of treatment encountered in literature was new features. Finally, it was observed that the standard A/B test was by far the most commonly used experiment design.

The mapping study \cite{auer2018current} by Auer and Felderer investigated the characteristics of the state of research on continuous experimentation. They observed that the intensity of research activities increased from year to year and that there is a high amount of collaboration between industry and academia. In addition, the authors observed that industrial and academic experts contributed equally to the field of continuous experimentation. Concerning the most influential publications (in terms of citations), the authors found that the most common research type among them is experience report. Another observation of the authors was that in total ten different terms were used for the concept of continuous experimentation.

To summarize, the two previous studies discussed continuous experimentation in terms of its applicability in industry sectors, the treatment types and experimentation designs reported, as well as the characteristics of the research in the field. In contrast to these two mapping studies, this study has a far more comprehensive synthesis. Furthermore, the two previous studies improved the rigor and completeness of the search and synthesis procedures.

\section{Research Method}
\label{sec:research_method}

Based on these two independently published systematic mapping studies~\cite{auer2018current,ros2018continuous2} we conducted a joint systematic literature review. Thus, the presented sets of papers from these two studies were used as starting sets. Forward snowballing was applied, by following the assumption from Wohlin~\cite{Wohlin2016} that publications acknowledge previous research. Relevant research publications were identified in the resulting sets. Next, the two sets were merged and the resulting set was studied to answer the respective research questions. Therefore, qualitative narrative syntheses~\cite{huang2018synthesizing} and a thematic synthesis~\cite{cruzes2011recommended} were conducted to answer the research question based on the found literature.

In the following, the research objective and the forward snowballing procedures are presented. Thereafter, the syntheses used to answer the research questions are described. 
Finally, the threats to validity are discussed.

\subsection{Research objective}

The aim of this research is to give an overview of the current state of knowledge about specific aspects of continuous experimentation. The research questions as stated in the introduction are on: 1) core constituents of a CE framework, 2) technical solutions within CE, 3) challenges with CE, and 4) benefits with CE. Based on the prior mapping studies we observed that there were many papers on models for processes and infrastructure, technical solutions, and challenges for CE and identified these as suitable targets for a systematic review.

\subsection{Forward snowballing} 
The two existing sets of papers emerging from the previous literature reviews \citer{auer2018current,ros2018continuous2}, were used as starting sets for forward snowballing. They were selected as starting sets, because both studies were in the field of continuous experimentation and they had similar research directions. Moreover, both studies were conducted within a short time of each other and had similar inclusion criteria. Hence, the authors are confident that the union of both selected paper sets is a good representation of the field of continuous experimentation in this context until 2017.

The forward snowballing was executed independently for each starting set. After having elaborated a protocol to follow, half of the authors worked on Set A (based on \citer{auer2018current}, with 82 papers) and half of them on the other Set B (based on \citer{ros2018continuous2}, with 62 papers).
In total, the starting sets contained 100 distinctive papers of which 44 papers were shared among both starting sets.
The citations were looked up on Google Scholar\footnote{\url{https://scholar.google.com/}}. Since the two previous mapping studies covered publications until 2017, the forward snowballing was conducted by considering papers within the time span 2017--2019. The snowballing was executed until no new publications were found.

In the process of snowballing, we used a joint set of inclusion and exclusion criteria. A paper was included if \emph{any} of the inclusion criteria applies, unless \emph{any} of the exclusion criteria applies. The decision was based primarily on the abstract of papers. If this was insufficient to make a decision, the full paper was examined. In doubt, the selection of a paper was discussed with at least one other author. The criteria were defined as such:
	
\paragraph{Inclusion criteria}
\begin{itemize}
	\item Any aspect of continuous experimentation (process, infrastructure, technical considerations, etc.)
	\item Any aspect of controlled experiments (designs, statistics, guidelines, etc.)
	\item Techniques that complement controlled experiments
\end{itemize}
\paragraph{Exclusion criteria}
\begin{itemize}
	\item Not written in English
	\item Not accessible in full-text
	\item Not peer reviewed or not a full paper
	\item Not a research paper: track, tutorial, workshop, talk, keynote, poster, book
	\item Duplicated study (the latest version is included)
	\item Primary focus on business-side of experimentation, advertisement, user interface, recommender system
\end{itemize}

The quality and validity of the included research publications were ensured through the inclusion and exclusion criteria. For instance, publications that did not go through a scientific peer-reviewing process were not considered according to the exclusion criteria. Moreover, to ensure that only mature work was included both vision papers with no evidence based contribution and short papers with preliminary results were excluded.

To summarize, the forward snowballing based on the starting Set A~\citer{auer2018current} resulted in 100 papers (Set A') and the starting Set B~\citer{ros2018continuous2} resulted in 88 papers (Set B'). After merging the two paper sets, a total of 128 distinctive papers represent the result of the applied forward snowballing.

\subsection{Synthesis}

To answer each research question, the collection of found papers was studied in more detail with respect to the individual research question. Therefore, two qualitative narrative syntheses~\citer{huang2018synthesizing} and one thematic synthesis~\cite{cruzes2011recommended} were conducted.

For the first two research questions a narrative synthesis was conducted for each question. This type of synthesis aggregates qualitative recurring themes within papers and provides a foundation for evidence-based interpretations of the themes in a narrative manner. 
Thus, the collected set of papers was studied under the heading of the two respective research questions (RQ1, RQ2) to identify relevant themes in it. Next, the found themes were summarized and identified patterns within them were reported. In addition, all papers were classified in terms of their \emph{research type} according to Wieringa et al.~\cite{wieringa2005requirements} to identify what solutions were applied (RQ2).
As a result, the findings represent an aggregated view on the components of a continuous experimentation framework (RQ1) and the technical solutions that are applied during experimentation (RQ2). The found components of a continuous experimentation framework are described in Section \ref{subsec:results_rq1}. An overview of the identified solutions can be found in Section~\ref{sec:results-rq2}.

\begin{figure}
    \centering
    \includegraphics[width=0.8\textwidth, trim=0cm 10.5cm 10cm 0cm,clip ]{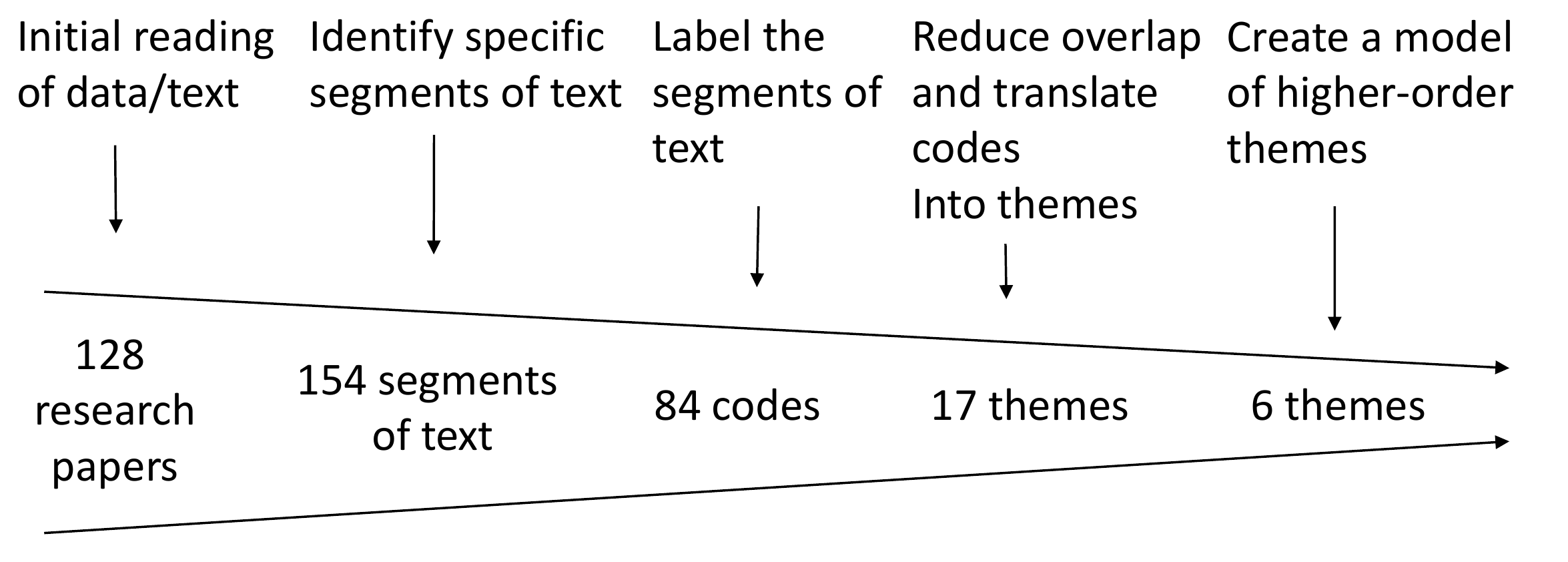}
    \caption{Applied thematic synthesis process (adapted from Cruzes et al. \cite{cruzes2011recommended,Cruzes14empirical})}
    \label{fig:thematic-process}
\end{figure}

For the third research question, a thematic synthesis following the steps and checklist proposed by Cruzes and Dyb\r{a} \citer{cruzes2011recommended} was conducted (see Fig.~\ref{fig:thematic-process}). In addition, the examples given in Cruzes et al.~\citer{Cruzes14empirical} were consulted. As an initial step, all 128 selected papers were read and in total 154 segments of text were identified. 
Next, each text segment was labeled with a code. A total of 84 codes were used to characterize the text segments. These codes were loosely based on terms that were identified in previous literature studies \citer{auer2018current,ros2018continuous2} and evolved during the labeling of the text segments. Thereafter, the codes that had overlapping themes were reduced into 17 themes. In the last step, these 17 themes were arranged according to 6 higher-order themes. The result of this analysis can be found in Section~\ref{sec:results:challenges}.

\begin{figure}
    \centering
    \includegraphics[width=0.8\textwidth, trim=0cm 7cm 0cm 0cm,clip ]{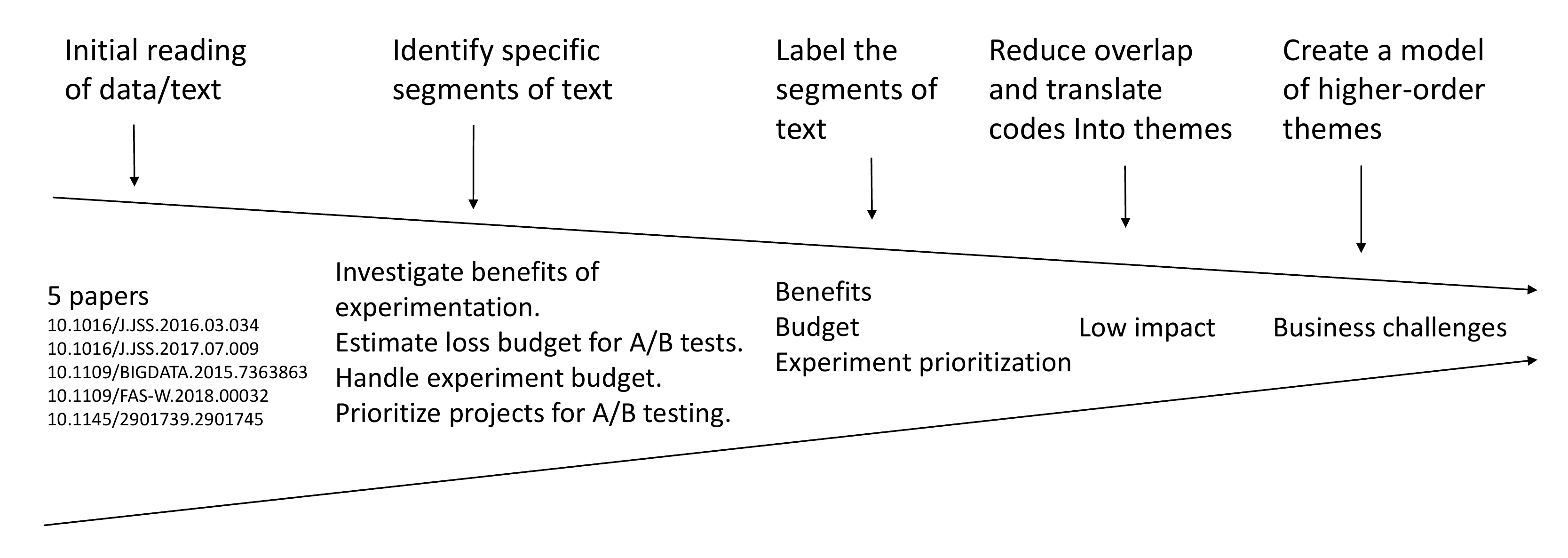}
    \caption{Schematic example of the thematic synthesis.}
    \label{fig:thematic-process-example}
\end{figure}

Fig.~\ref{fig:thematic-process-example} illustrates the thematic analysis process with the theme ``low impact''. Based on the reading of five papers, four text segments were extracted. These segments were labeled with the codes benefits, budget and experiment prioritization. In the next step, the common theme among the codes was identified and the codes were reduced to the theme ``low impact''. During the creation of the model of higher-order themes, this theme was assigned to the higher-order theme ``business challenges''. All text segments and codes can be found in the results of the study that are available online (see Section~\ref{subsec:threats-to-validity}).

\subsection{Threats to validity}\label{subsec:threats-to-validity}

In every step of this research possible threats to its validity were considered and minimized when possible. In the following, the potential threats are discussed to provide guidance in the interpretation of this work. This section is structured by the four criteria construct validity, internal validity, external validity and reliability by Easterbrook et al. \cite{Easterbrook_2008}.

\paragraph{Construct validity} This threat is about the validity of identification and selection of publications. A challenging threat to overcome is the completeness of the literature search without a biased view of the subject. To mitigate this threat, all papers from the start sets were used without any further exclusion. The larger start set (in comparison to applying the exclusion criteria from the start) was expected to lead to a broader coverage of the literature during the forward snowballing. Furthermore, the process of forward snowballing was adapted in the way that the candidate selection was tolerant about which papers to include, which increases the coverage of the literature search. However, publications may have been falsely excluded because of misjudgment. Nevertheless, we conducted two parallel forward snowballing searches by different authors based on slightly different starting sets, which should mitigate this threat. 

\paragraph{Internal validity} Threats that are caused by faulty conclusions could happen because of authors bias at the selection, synthesis of publications and interpretation of the findings. To mitigate this threat, a second author was consulted in case of any doubt. Nevertheless, activities like paper inclusion/exclusion and thematic synthesis inevitably suffer from subjective decisions. 

\paragraph{External validity} Threats to external validity covers to which extent the generalization of the results is justified. As the aim of this study is to give an overview of continuous experimentation and to explore the future work items in continuous experimentation, the results should not be generalized beyond continuous experimentation. Therefore, this threat is neglectable.

\paragraph{Reliability} This threat focuses on the reproducibility of the study and the results. To mitigate this threat every step and decision of the study were recorded carefully and the most important decisions are reported. The results of the study are available online~\cite{auer_ros_kaltenbrunner_runeson_felderer_2021}. This enables other researchers to validate the decision made on the data. Furthermore, it allows to repeat the study. 

\section{Results}
\label{sec:results}

In this section the results of the literature review are presented according to the research questions.

\subsection{What are the core constituents of a CE framework (RQ1)?}\label{subsec:results_rq1}

To conduct continuous experimentation, an organization has to have some constituents of a framework for experimentation. There is some process involved (implicit or explicit) and some infrastructure is required, which includes a toolchain as well as organizational processes. In the following, both aspects of an experiment, the process and its supporting infrastructure are discussed in detail.

\subsubsection{Experiment process}

The experiment process can be described in a model that gives a holistic view of the phases and environment around experimentation. Most studies on experiment processes present qualitative models based on interview data. Two models describe the overall process of experimentation. First, the reference model RIGHT (Rapid Iterative value creation Gained through High-frequency Testing) by Fagerholm et al.~\citer{fagerholm2017right} contains both an infrastructure architecture and a process model for continuous experimentation. The process model builds on the Build, Measure, Learn~\cite{ries2011lean} cycle of Lean Startup. The process in Figure~\ref{fig:ce_process} is a simplified view of RIGHT. Second, the HYPEX (Hypothesis Experiment Data-Driven Development) model is another earlier process model by Holmstr{\"o}m Olsson and Bosch~\citer{olsson2014from}. In comparison to the RIGHT model, it is less complete in scope, however it does go into further details in hypothesis prioritization using a gap analysis.

Kevic et al.~\citer{kevic2017characterizing} present concrete numbers on the experiment process used at Microsoft Bing through a source code analysis. They have three main findings: 1) code associated with an experiment is larger in terms of files in a changeset, number of lines, and number of contributors; 2) experiments are conducted in a sequence of experiments lasting on average 42 days, where each experiment is on average conducted for one week; and 3) only a third of such sequences are eventually shipped to users.

In addition to the general models described above, several models deal with a specific part of the experiment cycle. 
The CTP (Customer Touchpoint) model by Sauvola et al.~\citer{sauvola2018continuous} focuses on user collaboration and describes the various ways that user feedback can be involved in the experimentation stages. 
Amatrian~\citer{amatriain2013beyond} and Gomez-Uribe and Hunt~\citer{uribe2015netflix} describe their process for experimentation on their recommendation system at Netflix, in particular how they use offline simulation studies with online controlled experiments.
In the ExG Model (Experimentation Growth), by Fabijan et al.~\citer{fabijan2017evolution,fabijan2018online2}, organizations can quantitatively gauge their experimentation on technical, organizational, and business aspects.
In another model by Fabijan et al.~\citer{fabijan2018effectivePAPER} they describe the process of analyzing the results of experiments and present a tool that can make the process more effective, by e.g. segmenting the participants automatically and highlighting the presence of outliers. 
Finally, Mattos et al.~\citer{mattos2018activity} present a model that discuss details on activities and metrics on experiments.

\subsubsection{Infrastructure}

Depending on what type of experimentation is conducted, different infrastructure is required. For controlled experimentation, in particular, technical infrastructure in the form of an \emph{experimentation platform} is critical to increase the scale of experimentation. At the bare minimum it needs to divide users into experiment groups and report statistics. Gupta et al.~\citer{gupta2018anatomy} at Microsoft have detailed the additional functionality of their experimentation platform. Also Schermann et al.~\citer{schermann2018doing} have described attributes of system and software architecture suitable for experimentation, namely, that micro-service-based architectures seem to be favored. Some experimentation platforms are specialized to specific needs: automation~\citer{mattos2017your}, or describing deployment through formal models~\citer{schermann2016bifrost}, or how experimentation can be supported by non-software engineers~\citer{koukouvis2016ab,firmenich2018usability}.

There are also non-technical infrastructure requirements, regardless of the type of experimentation in use. The required roles are~\citer{fitzgerald2017continuous}: data scientists, release engineers, user researchers, and the standard software engineering roles. Also, an organizational culture~\citer{kohavi2009online,xu2015from} that is open towards experimentation is needed. For example, Kohavi et al.~\citer{kohavi2009online} explain that managers can hinder experimentation if they overrule results with their opinions. They call the phenomenon the highest paid persons opinion (HiPPO).

While experimentation is typically associated with large companies, like Microsoft or Facebook, there are three interview studies that discuss experimentation at startups specifically~\citer{bjorklund2013lean,gutbrod2017how,fagerholm2017right}. As argued by Gutbrod et al.~\citer{gutbrod2017how}, startup companies often guess or estimate roughly about the problems and customers they are addressing. Thus, there is a need for startup companies to be more involved with experimentation, although they have less infrastructure in place.

Finally, we would like to call attention to some of the few case studies and experience reports on experimentation on ``ordinary'' software companies, which are neither multi-national corporations nor startups~\citer{ros2018continuous,rissanen2015continuous,yaman2018continuous}; in e-commerce, customer relations, and gaming industry respectively. None of these papers are focused on infrastructure, but do mention that infrastructure needs to be implemented. Risannen et al.~\citer{rissanen2015continuous} mentions additional challenges when infrastructure must be implemented on top of a mature software product. In summary, this indicates that infrastructure requirements are modest unless scaling up to multi-national corporation levels with millions of users.

\subsection{What technical solutions are applied in what phase within CE (RQ2)}\label{sec:results-rq2}

\begin{figure}
\centering
\small
\input{solutions.tex}
\caption{Solutions applied within continuous experimentation arranged by main phase of experimentation from Section~\ref{sec:background_process}.}
\label{fig:rq2-overview}
\end{figure}
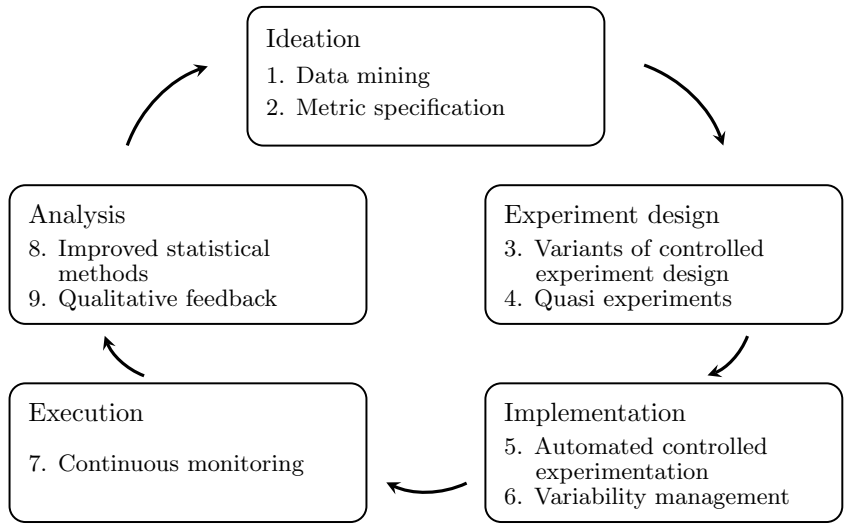

The study of the selected publications revealed many different types of solutions that were summarized by common themes. Figure \ref{fig:rq2-overview} gives an overview of the identified solutions organized in the phases of experimentation in Figure~\ref{fig:ce_process}. 

\subsubsection{Data mining}

Data from previous experiments can be used to make predictions or mine insights to either improve the reliability of the experiment or for ideation. There were three specific solutions for data mining in continuous experimentation: 1)~calculating variance of metrics through bigger data sets than just one experiment at Netflix~\citer{xie2016improving}, Microsoft~\citer{deng2015objective,deng2013improving}, Google~\citer{hohnhold2015focusing}, and Oath~\citer{appiktala2017demystifying}; 2)~mining for invalid tests through automatic diagnosis rules at LinkedIn~\citer{chen2019how} and Sevenval Technologies~\citer{nolting2016context}; and finally 3)~to extract insights from segments of the users, by detecting if a treatment is more suitable for those specific circumstances~\citer{duivesteijn2017have}, this technique is applied at Snap~\citer{xie2018false} and Microsoft~\citer{fabijan2018effectivePAPER}.

\subsubsection{Metric specification}
Defining measurements for software is difficult. At Microsoft they have hundreds of metrics in place for each experiment, they recommend organizing metrics in a hierarchy~\citer{machmouchi2016principles} and evaluating how well metrics work~\citer{deng2016data,dmitriev2016measuring}. At Yandex, they pair OEC metrics with a statistical significance test to create an overall acceptance criteria (OAC) instead~\citer{drutsa2015practical}. Several pieces of work are on defining and improving usability metrics, especially from Yandex~\citer{budylin2018consistent,drutsa2017using,kharitonov2017learning}. Also at Microsoft they have a rule-based classifier where each user action is either a frustration or benefit signal~\citer{machmouchi2017beyond}.

Some general guidelines for defining metrics follow. At Microsoft~\citer{deng2016data,dmitriev2016measuring,machmouchi2016principles}, they have hundreds of metrics for each experiment (in addition to a few OEC). Machmouchi and Buscher~\citer{machmouchi2016principles} from Microsoft describe how their metrics are interpreted in a hierarchy in their tool (similar to Fabijan et al.~\citer{fabijan2018effectivePAPER} also at Microsoft). At the top of the hierarchy are statistically robust metrics (meaning they tend not to give false positives) and at the bottom are feature specific metrics that are allowed to be more sensitive. They have also developed methods to evaluate how well metrics work. Dmitriev et al.~\citer{dmitriev2016measuring} give an experience report on how metrics are evaluated at Microsoft system in practice. Deng et al.~\citer{deng2016data} define metrics for evaluating metrics: directionality and sensitivity. They measure respectively whether a change in the metric aligns with good user experience and how often it detects a change in user experience.

Usability metrics are hard to define since they are not directly measurable without specialized equipment, such as eye-tracking hardware or focus groups. The measurements that are available, such as clicks or time spent on the site, do not directly inform on whether a change is an improvement or degradation in user experience. In addition, good user experience does not necessarily correlate positively with business value, e.g. clickbait titles for news articles are bad user experience but generate short term revenue. Researchers from Yandex~\citer{budylin2018consistent,drutsa2015sign,drutsa2015future,drutsa2017using,kharitonov2017learning,poyarkov2016boosted} are active in this area, with the following methods focused on usability metrics: detecting whether a change in a metric is a positive or negative user experience~\citer{drutsa2015sign}; learning sensitive combinations of metrics~\citer{kharitonov2017learning}; quantifying and detecting trends in user learning~\citer{drutsa2017using}; predicting future behavior to improve sensitivity~\citer{drutsa2015future}; applying machine learning for variance reduction~\citer{poyarkov2016boosted}; and finally correcting misspecified usability metrics~\citer{budylin2018consistent}. Machmouchi et al.~\citer{machmouchi2017beyond}, at Microsoft, designed a rule-based classifier where each user action is either a frustration or benefit signal; the tool then aggregates all such user actions taken during a session into a single robust utility metric.

\subsubsection{Variants of controlled experiments design}

Most documented experiments conducted in industry are univariate A/B/n-tests~\citer{ros2018continuous2}, where one or more treatments are tested against a control. Extensions to classical designs include a two-staged approach to A/B/n tests~\citer{deng2014statistical} and a design to estimate causal effects between variables in a multivariate test (MVT)~\citer{peysakhovich2018learning}. MVTs are cautioned against~\citer{kohavi2014seven} because of their added complexity. In contrast, other researchers take an optimization approach using lots (see Section~\ref{subsec:automated_controlled}) of variables with multi-armed bandits~\citer{claeys2017regression,hill2017efficient,mattos2018optimization,ros2018continuous} or search-based methods ~\citer{miikkulainen2017conversion,ros2017automated,tamburrelli2014towards}. Also mixed methods research is used to combine quantitative and qualitative data. Controlled experiments require deployment, feedback from users at earlier stages of development can thus be cheaper. There are works on combining results of such qualitative methods~\citer{bosch2016speed} and collecting it in parallel with A/B tests~\citer{speicher2014ensuring}.

\subsubsection{Quasi-experiments}

A quasi-experiment (or natural experiment) is an experiment that is done sequentially instead of in parallel; this definition is the same as in empirical research in general~\cite{wohlin2012experimentation}. The reason for doing it is that it has a lower technical complexity. In fact, any software deployment can have its impact measured by observing the effect before and after deployment. The drawback of this is that analyzing the results can be difficult due to the high risk of having external changes affect the result. That is, if anything extraordinary happens roughly at the same time as the release it might not be possible to properly isolate the results. Since the world of software is in constant change the use of quasi-experiments is challenging. The research directions on quasi-experiments involve how to eliminate external sources of noise to get more reliable results. This is studied at Amazon~\citer{hill2015measuring} and Linkedin~\citer{xu2016evaluating}, particularly for environments were control is continuous deployment is hard (such as mobile app development).

\subsubsection{Automated controlled experimentation with optimization algorithms}
\label{subsec:automated_controlled}
With an optimization approach, the allocation of users to the treatment groups is dynamically varied to optimize an OEC, such that treatments that perform well continuously receive more and more traffic over time. With sufficient automation, these techniques can be applied to lots of treatment variables simultaneously. This is not a replacement for classical designs; in an interview study by Ros and Bjarnason~\citer{ros2018continuous}, they explain that such techniques are often validated themselves using A/B tests. In addition, based on the studies included here, only certain parameters are eligible, such as the design and layout of components in a GUI, or parameters to machine learning algorithms or recommender systems. Some of these optimizations are black-box methods, where multiple variables are changed simultaneously and with little opportunity to make statistical inferences from the experiments.

Tamburelli and Margara~\citer{tamburrelli2014towards} proposed search-based methods (i.e. genetic algorithms) for optimization of software, and Iitsuka and Matsuo~\citer{iitsuka2015website} demonstrated a local search method with a proof of concept on web sites. Miikkulainen~\citer{miikkulainen2017conversion}, at Sentient Technologies, have a commercial genetic algorithm profiled for optimizing e-commerce web sites. 
Bandit optimization algorithms are also used in industry at Amazon~\citer{hill2017efficient} and AB Tasty~\citer{claeys2017regression}, it is a more rigorous formalism that requires the specification of a statistical model on how the OEC behaves. Ros et al.~\citer{ros2017automated} suggested a unified approach of genetic algorithms and bandit optimization.
Similar algorithms exist to handle continuous variables, as is needed for hardware parameters~\citer{gerostathopoulos2018tool,mattos2018optimization} and for optimizing machine learning and compiler parameters~\citer{letham2019constrained}.

Two studies apply optimization~\citer{kharitonov2015optimised,schermann2018search} to scheduling multiple standard A/B tests to users, where only a single treatment is administered to each user. The idea is to optimize an OEC without sacrificing statistical inference.

\subsubsection{Variability management}

Experimentation incurs increased variability\textemdash by design\textemdash in a software system. This topic deals with solutions in the form of tools and techniques to manage said variability. In terms of an experiment platform, this can be part of the \emph{experiment execution service} and/or the \emph{experimentation portal}~\citer{gupta2018anatomy}.

There have been attempts at imposing systematic constraints and structure in the configuration of how the variables under experimentation interact with formal methods. C{\'a}mara and Kobsa~\citer{cmara2009facilitating} suggest using a feature model of the software parameters in all experiments. This work has not advanced beyond a proof-of-concept stage. 

Neither in our study, nor in the survey by Schermann et al.~\citer{schermann2018doing}, is there any evidence of formal methods in a dynamic and constantly changing experimentation environment. The focus of the tools in actual use are rather on flexibility and robustness~\citer{bakshy2014designing,tang2010overlapping}. 
Rahman et al.~\citer{rahman2016feature} studied how feature toggles are used in industry. Feature toggles are ways of enabling and disabling features after deployment, as such they can be used to implement A/B testing. They were found to be efficient and easy to manage but adds technical debt.

A middle ground between formal methods and total flexibility has evolved in the tools employed in practice. Google has proprietary tools in place to manage overlapping experiments in large scale~\citer{tang2010overlapping}. In their tools, each experiment can claim resources used during experimentation and a scheduler ensures that experiments can run in parallel without interference. Facebook has published an open-source framework (PlanOut) specialized for configuring and managing experiments~\citer{bakshy2014designing}, it features a namespace management system for experiments running iteratively and in parallel. SAP has a domain-specific language~\citer{westermann2013experiment} for configuring experiments that aims at increasing automation.
Finally, Microsoft has the ExP platform, but none of the selected papers focus solely on the variability management aspect of it.

\subsubsection{Improved statistical methods}

The challenges with experimentation motivate improved statistical techniques specialized for A/B testing. There are many techniques for fixing specific biases, sources of noise, etc: a specialized test for count data at SweetIM~\citer{borodovsky2011ab}; fixing errors with dependent data at Facebook~\citer{bakshy2013uncertainty}; improvements from the capabilities of A/A testing on diagnosis (which tests control vs control expecting no effect) at Yahoo~\citer{zhao2016online} and Oath~\citer{chen2017faster}; better calculation of overall effect for features with low coverage at Microsoft~\citer{deng2015diluted}; fixing errors from personalization interference at Yahoo~\citer{das2013when}; fixing tests under telemetry loss at Microsoft~\citer{gupchup2018trustworthy}; correcting for selection bias at Airbnb~\citer{lee2018winners}; and algorithms for improved gradual ramp-up at Google~\citer{medina2018online} and LinkedIn~\citer{xu2018sqr}.

\subsubsection{Continuous monitoring}

Aborting controlled experiments pre-maturely in case of outstanding or poor results is a hotly debated topic on the internet and in academia, under the name of continuous monitoring, early stopping, or continuous testing. The reason for wanting to stop early is to reduce opportunity costs and to increase development speed. It is studied by Microsoft~\citer{deng2016continuous}, Yandex~\citer{kharitonov2015sequential}, Optimizely~\citer{johari2017peeking}, Walmart~\citer{abhishek2017nonparametric}, and Etsy~\citer{ju2019sequential}.  This concept is similar to the continuous monitoring used by researchers in the DevOps community and continuous software engineering~\citer{fitzgerald2017continuous} where it refers to the practice of monitoring a software system and sending alerts in case of faults. The issue with continuous monitoring of experiments is the increased chance of getting wrong results if carried out incorrectly.
Traditionally, the sample size of an experiment is defined beforehand through a power calculation. If the experiment is continuously monitored with no adjustment, then the results will be skewed with inflated false negative and positive error rates.

\subsubsection{Qualitative feedback}
\label{sec:results_qualitative}
While the search strategy in this work was focused on controlled experiments, research on qualitative feedback was also included from experience reports on using many different types of feedback collecting methods, for example at Intuit~\citer{bosch2012building,bosch2016speed} and Facebook~\citer{feitelson2013development}. The qualitative methods are used as complements to quantitative methods, either as a way to better explain results or as a way to obtain feedback earlier in the process, before a full implementation is built. That is, qualitative feedback can be collected on early user experience sketches or mock-ups. Another use of qualitative methods is to elicit hypotheses that can be used as a starting point for an experiment. Examples of methods include focus groups, interviews, and user observations. 

In addition, at Unister~\citer{speicher2014ensuring} the authors explain how they collect qualitative user feedback in parallel with A/B tests, such that the feedback is split by experiment group. According to the authors, this seems to be a way to get the best of both quantitative and qualitative worlds. It does require implementing a user interface for collecting the feedback in a non-intrusive way in the product. Also, the qualitative feedback will not be of as high quality as when it is done in person with e.g. user observation or focus groups.

\subsection{What are the challenges with continuous experimentation (RQ3)?}
\label{sec:results:challenges}

Continuous experimentation encompasses a lot of the software engineering process, it requires both infrastructure support and a rigorous experimentation process that connects the software product with business value. As such, many things can go wrong and the challenges presented here is an attempt at describing such instances. Most of the research on challenges is evaluation research, with interviews or experience reports. 
Many of the challenges are severe, in that they present a hurdle that must be overcome to conduct continuous experimentation. A failure in any of the respective category of challenges will make an experiment: unfeasible due to technical reasons, not considered by unresponsive management, untrustworthy due to faulty use of statistics, or without a business case. 
The analysis of the papers revealed six categories of challenges (see Table \ref{tab:challenges}) that are discussed in the following in more detail.

\begin{table}[tpbh]
    \centering
    \renewcommand{\arraystretch}{1.2}
    \caption{Summary of challenges with continuous experimentation per category with description and key references that focus on them.}
    \label{tab:challenges}
    \footnotesize
\vspace{-15pt} 
\begin{tabularx}{\textwidth}{@{}>{\raggedright}p{7.5em}Xr@{}}
        \toprule
        \textbf{Challenge} & \textbf{Description} & \textbf{References} \\
        
        \midrule
        \multicolumn{2}{l}{\emph{1. Cultural, organizational, and managerial challenges}} \vspace{-2pt}\\
        \midrule
        
        Knowledge building & There are many roles and skills required, so staff need continuous training.  & \citer{kohavi2013online,rissanen2015continuous,yaman2017introducing} \\
        Micromanagement & Experimentation requires management to focus on the process (c.f. HiPPO in Section \ref{sec:challenge_soft}). & \citer{kohavi2009online} \\
        Lack of adaption & Engineers need to be onboarded on the process as well as managers. & \citer{lindgren2016raising} \\
        Lack of communication & Departments and teams should share their results to aid each other. & \citer{rissanen2015continuous,yaman2017introducing} \\
        
        \midrule
        \multicolumn{2}{l}{\emph{2. Business challenges}} \vspace{-2pt}\\
        \midrule
        
        Low impact & Experimentation might focus efforts on incremental development with insufficient impact.  & \citer{fitzgerald2017continuous,olsson2017experimentation} \\
        Relevant metrics & The business model of a company might not facilitate easy measurement. & \citer{dmitriev2017dirty,fabijan2018online2,lindgren2016raising} \\
        Data leakage & Companies expose internal details about their product development with experimentation. & \citer{conti2018spot} \\
        \midrule
        \multicolumn{2}{l}{\emph{3. Technical challenges}} \vspace{-2pt}\\
        \midrule
        
        Continuous delivery & The CI/CD pipeline should be efficient to obtain feedback fast. & \citer{fabijan2018online2,lindgren2016raising,schermann2018doing} \\
        Continuous deployment & Obstacles exists to putting deliveries in production, e.g. on-premise installations in B2B. & \citer{rissanen2015continuous}\\
        Experimental control & Dividing users into experimental groups have many subtle failure possibilities. & \citer{crook2009seven,dmitriev2016pitfalls,kohavi2008controlled} \\
        
        \midrule
        \multicolumn{2}{l}{\emph{4. Statistical challenges}} \vspace{-2pt}\\
        \midrule
        Exogenous effects & Changes in environment can impact experiment results, e.g. trend effects in fashion. &  \citer{dmitriev2016pitfalls,kohavi2012trustworthy}\\
        Endogenous effects & Experimentation itself causes effects, such as carry-over or novelty effects. & \citer{dmitriev2016pitfalls,lu2014separation} \\
        
        \midrule
        \multicolumn{2}{l}{\emph{5. Ethical challenges}} \vspace{-2pt}\\
        \midrule
        Data privacy & GDPR gives users extensive rights to their data which companies must comply with. & \citer{yaman2017notifying} \\
        
        Dark patterns & A narrow focus on numbers only can lead to misleading user interfaces. & \citer{jiang2019whos} \\
        
        \midrule
        \multicolumn{2}{l}{\emph{6. Domain specific challenges}} \vspace{-2pt}\\
        \midrule
        Mobile & The app marketplaces impose constraints on deployment and variability. & \citer{lettner2013enabling,xu2016evaluating,yaman2018continuous} \\
        Cyber-physical systems & Making continuous deployments can be infeasible for cyber-physical systems. & \citer{bosch2016data,giaimo2017considerations,mattos2018challenges}  \\
        Social media & Users of social media influence each other which impacts the validity of experiments. & \citer{azevedo2018estimatino,backstrom2011network,choi2017estimation} \\
        E-commerce & Experimentation needs to be able to differentiate effects from products and software changes. & \citer{goswami2015controlled,wang2018designing} \\
        \bottomrule
\end{tabularx}
\end{table}

\subsubsection{Cultural, organizational, and managerial challenges}
\label{sec:challenge_soft}
The challenges to organizations and management are broad in scope, including: difficulty in changing the organizational culture to embrace experimentation~\citer{lindgren2016raising}; building experimentation skills among employees across the whole organization~\citer{kohavi2013online,yaman2017introducing}; and finally communicating results and coordinating experiments in business to business, where there are stakeholders involved across multiple organizations~\citer{rissanen2015continuous,yaman2017introducing}.

A fundamental challenge that has to be faced by organizations adopting continuous experimentation, is the shift from the highest-paid person's opinion (HiPPO)~\citer{kohavi2007practical,kohavi2009online} to data-driven decision making. If managers are used to making decisions about the product then they might not take account of experimental results that might run counter to their intuition. Thus, decision-makers must be open to have their opinions changed by data, else the whole endeavor with experimentation is useless.

\subsubsection{Business challenges}
The premise behind continuous experimentation is to increase the business value of software development efforts. The most frequent challenge in realizing this is defining relevant metrics that measure business value~\citer{dmitriev2016pitfalls,dmitriev2017dirty,fabijan2018online2,lindgren2016raising,yaman2017introducing}. In some instances the metric is only indirectly connected to business, for example in a business-to-business (B2B) company with a revenue model that is not affected by the software product, then improving product quality and user experience will not have a direct business impact. Also, the impact of experiments might not be sufficient in terms of actual effect~\citer{kohavi2014seven,olsson2017experimentation}. Fitzgerald and Stol~\citer{fitzgerald2017continuous} argue that continuous experimentation and innovation can lead to incremental improvements only, at the expense of more innovative changes that could have had a bigger impact. Another business challenge of continuous experimentation was highlighted by Conti et al.~\citer{conti2018spot}; they crawled web sites repeatedly and tried to automatically detect a difference in server responses. Thereby they showed how easily such data leakage can facilitate industrial espionage on what competitors are developing.

\subsubsection{Technical challenges}
Efficient continuous deployment facilitates efficient experimentation. Faster deployment speed shortens the delay between a hypothesis and the result of an experiment. The ability to have an efficient continuous delivery cycle is cited as a challenge both for large~\citer{kohavi2008controlled} and small companies~\citer{fabijan2018online2,lindgren2016raising,schermann2018doing}. In addition, continuous deployment is further complicated in companies involved in business to business (B2B)~\citer{rissanen2015continuous}, where deployment has multiple stakeholders involved over multiple organizations.

In a laboratory experiment setting, it is possible to control variables such as ensuring homogeneous computer equipment for all groups and ensuring that all groups have equal distribution in terms of gender, age, education, etc. For online experiments, such controls are much harder due to subtle technical reasons. Examples therefore are: users assigned incorrectly to groups due to various bugs~\citer{kohavi2008controlled}; users changing groups because they delete their browsing history or multiple persons share the same computer~\citer{coey2016people,deng2017trustworthy,dmitriev2016pitfalls,kohavi2008controlled}; and robots from search engines cause abnormal traffic affecting the results~\citer{crook2009seven,kohavi2011unexpected}.

\subsubsection{Statistical challenges}

Classical experimental design as advocated by the early work on continuous experimentation and A/B-testing~\citer{kohavi2007practical} does not account for time series. Not only can it be hard to detect the presence of effects related to trends, but they can also have an effect on the results. Some of these trend effects occur due to outside influence, so-called \emph{exogenous effects}, for example, due to seasonality caused by fashion or other events which can affect traffic~\citer{dmitriev2016pitfalls,kohavi2012trustworthy}. With domain knowledge, these effects can be accounted for. For example in e-commerce, experiment results obtained during Christmas shopping week might not transfer to the following weeks. 

Other statistical challenges are caused by the experimentation itself, called \emph{endogenous effects}, such as the carryover effect~\citer{kohavi2012trustworthy,lu2014separation} where the result of an experiment can affect the result of a following experiment. There are also endogenous effects caused intentionally, through what is known as ramp up, where the traffic to the test group is initially low (such as 5\%/95\%) and incrementally increased to the full 50\%/50\% split. This is done to minimize the opportunity cost of a faulty experiment design. It can be difficult to analyze the results of such experiments~\citer{crook2009seven,kohavi2011unexpected}. Furthermore, learning and novelty effects where the users change their impression of the feature after using it for a while are challenging~\citer{dmitriev2016pitfalls,lu2014separation}.

Endogenous effects will be hard to foresee until experimentation is implemented in a company. As such, handling statistical challenges is an ongoing process that will require more and more attention as experimentation is scaled up.

\subsubsection{Ethical challenges}
Whenever user data is involved there is a potential for ethical dilemmas. 
When Yaman et al.~\citer{yaman2017notifying} surveyed software engineering practitioners, the only question they agreed on was that users should be notified if personal information is collected. Since GDPR went into effect in 2018 this is now a requirement.
Jian et al.~\citer{jiang2019whos} investigate how A/B testing tools are used in illegal discrimination for certain demographics, e.g., by adjusting prices or filtering job ads. These are examples of what is known as \emph{dark patterns} in the user experience (UX) research community~\cite{gray2018dark}. The study was limited to sites using front end Optimizely (a commercial experimentation platform) metadata. 

\subsubsection{Domain specific challenges}
Some software sectors have domain-specific challenges or techniques required for experimentation, of which in the analysis of the papers four prominent domains were found: 1) mobile apps, 2) cyber-physical systems, 3) social media, and 4) e-commerce. Whether or not all of these concerns are domain-specific or not is debatable. However, these studies were all clear on what domain their challenges occurred in.

There is a bottleneck in continuous deployment to the proprietary application servers of Android Play or Apple's App Store, which imposes a bottleneck on experimentation for mobile apps. Lettner et al.~\citer{lettner2013enabling} and Adinata and Liem~\citer{adinata2014ab} have developed libraries that load new user interfaces at run time, which would otherwise (at the time of writing in year 2013 and 2014 respectively) require a new deployment on Android Play. Xu et al.~\citer{xu2016evaluating} at LinkedIn instead advocate the use of quasi-experimental designs. Finally, Yaman et al.~\citer{yaman2018continuous} have done an interview study on continuous experimentation, where they emphasize user feedback in the earlier stages of development (that do not require deployment).

Embedded systems, cyber-physical systems, and smart systems face similar challenges to mobile apps, namely continuous deployment. None of the studied publications of this study claims wide\-spread adoption of experimentation at an organizational level. This suggests that research of experimentation for embedded software is in an early stage. 
Mattos et al.~\citer{mattos2018challenges} and Bosch and Holmstr{\"o}m Olsson~\citer{bosch2016data} outline challenges and research opportunities in this domain, among them are: continuous deployment, metric definition, and privacy concerns. Bosch and Eklund~\citer{bosch2012eternal,eklund2012architecture} describe required architecture for experimentation in this domain with a proof-of-concept on vehicle entertainment systems. Giaimo et al.~\citer{giaimo2017considerations,giaimo2016continuous} cite safety concerns and resource constraints for the lack of continuous experimentation.

The cyber-physical systems domain also includes experimentation where the source of noise is not human users, but rather hardware. The research on self-adaptive systems overlap with continuous experimentation: Gerostathopoulos et al.~\citer{gerostathopoulos2016architectural} have described an architecture for how self-adaptive systems can perform experimentation, with optimization algorithms~\citer{gerostathopoulos2018adapting} that can handle non-linear interactions between hardware parameters~\citer{gerostathopoulos2018cost}. In addition, two pieces of work~\citer{buchert2015survey,jayasinghe2013automated} on distributed systems focus on experimentation, with a survey and a tool on how distributed computing can support experimentation for e.g. cloud providers.

Backstrom et al.~\citer{backstrom2011network} from Facebook describe that users of social media influence each other across experiment groups (thus violating the independence assumption of statistical tests); they call it the network effect. It is also present at Yahoo~\citer{katzir2012framework} and LinkedIn~\citer{gui2015network,saveski2017detecting,xu2015from}. The research on the network effect includes: ways of detecting it~\citer{saveski2017detecting}, estimating its effect on cliques in the graph~\citer{azevedo2018estimatino,choi2017estimation}, and reducing the interference caused from it~\citer{eckles2016design}.

The final domain considerations come from e-commerce. At Walmart, Goswami et al.~\citer{goswami2015controlled} describe the challenges caused by seasonality effects during holidays and how they strive to minimize the opportunity cost caused by experimentation. At Ebay, according to Wang et al.~\citer{wang2018designing}, the challenges are caused by the large number of auctions that they need to group with machine learning techniques for the purpose of experimental control.

\subsection{What are the benefits with continuous experimentation (RQ4)?}

Many authors mention the benefits of CE only in passing as motivation~\citer{bosch2012building,kohavi2009online}, few papers explicitly mention them (e.g. \citer{fabijan2017benefits}).

Bosch~\citer{bosch2012building} mentions the reduced cost of collecting passive customer feedback with continuous experimentation in comparison with active techniques like surveys. Also, Bosch claims that customers have come to expect software services to continuously improve themselves and that experimentation can provide the means to do that in a process that can be visible to users.
Kohavi et al.~\citer{kohavi2009online} claim that edge cases that are only relevant for a small subset of users can take a disproportionate amount of the development time. Experimentation is argued for as a way to focus development, by first ensuring that a feature solves a real need with a small experiment and then optimizing the respective feature for the edge cases with iterative improvement experiments. In this way, unnecessary development on edge cases can be avoided if a feature is discarded early on.

Fabijan et al.~\citer{fabijan2017benefits} focus solely on benefits, differentiated between three levels as follows. 1) In the \emph{portfolio level}, the impact of changes on the customer as well as business value can be measured which is of great benefit to company-wide product portfolio development. 2) In the \emph{product level}, the product receives incrementally improvement quality and reduced complexity by removing unnecessary features. 
Finally, 3) in the \emph{team level of benefits}, the findings of experiments support the related teams to prioritize their development activities given the lessons learned from the conducted experiments. Another benefit for teams with continuous experimentation is that team goals can be expressed in terms of metric changes and their progress is measurable. 

\section{Discussion}
\label{sec:discussion}

This study builds on two prior independent mapping studies to provide an overview of the conducted research. This review has been conducted to answer four research questions that can guide practitioners. In the following, the results of the study are discussed for each research question, in the form of recommendations to practitioners and implications for researchers.

\subsection{Required frameworks (RQ1)}

The first research question (RQ1) about the core constituents of a framework for continuous experimentation revealed two integral parts of experimentation, the \emph{experimentation process} and the technical as well as organizational \emph{infrastructure}.

\subsubsection{Process for continuous experimentation}
In the literature, several experimentation process models were found on the phases of conducting online controlled experimentation. They describe the overall process~\citer{fagerholm2017right}, represent the established experiment process of organizations~\citer{kevic2017characterizing}, or cover specific parts of the experiment cycle~\citer{fabijan2018effectivePAPER}. Given that all models describe a process with the same overall objective of experimentation, it can become difficult to decide between them. Two reference models are published \citer{fagerholm2017right,olsson2014from}, which may be used as a basis for future standardization of the field. Future research is needed to give guidance in the selection between models and variants.

Many of the experience reports~\citer{kohavi2014seven,kohavi2011unexpected} warn about making experiments with too broad scope, instead they recommend that all experiments should be done on a minimum viable product or feature~\citer{fagerholm2017right}. However, the warnings all come from real lessons learned caused by having done such expensive experiments. We believe that the current process models do not put sufficient emphasis on conducting small experiments. For example, they could make a distinction between prototype experiments and controlled experiments on a completed project. That way if the prototype reveals flaws in the design it avoids a full implementation.

As such, our recommendation to practitioners in regards to process is to follow one of the reference experimentation processes~\citer{fagerholm2017right,olsson2014from} and in addition add the following two steps to minimize the cost of experiments. First, to spend more time before experimentation to ensure that experiments are really on a minimum viable feature by being diligent about what requirements are strictly needed at the time. Second, that experiments should be pre-validated with prototypes, data analysis, etc.


\subsubsection{Infrastructure for continuous experimentation}
The research on the infrastructure required to enable continuous experimentation was primarily focused on large scale applications within mature organizations (e.g. Microsoft \citer{gupta2018anatomy}). One reason for this focus may be the large number of publications (e.g. experience reports) from researchers associated with large organizations. The large number of industrial authors indicates a high interest of practitioners in the topic.
However, it should not restrict the community's focus on large scale applications only. The application of continuous experimentation within smaller organizations has many open research questions. These organizations provide additional challenges on experimentation because of their probably less already existing infrastructure and smaller user base. For example, the development of sophisticated experimentation platforms may not be feasible in the extent to which it is for large organizations. Thus, lightweight approaches to experimentation that do not require large up-front investments could make experimentation more accessible to smaller organizations.

Technical infrastructure has not been reported as being a significant hurdle for any of the organizations in which continuous experimentation was introduced in this study. The technical challenges seem to appear later on when the continuous experimentation process has matured and the scale of experimentation needs to ramp up. Rather, the organizational infrastructure seems to be what might cause an inability to conduct experimentation. The challenges presented in Section~\ref{sec:results:challenges} support this claim too, so the more severe infrastructural requirements appear to be organizational~\citer{fitzgerald2017continuous} and culture oriented~\citer{kohavi2009online,xu2015from}, at least to get started with experimentation. The reason for this is that experimentation often involves decision making that traditionally fall outside the software development organization. For example, deciding on what metric software should be optimized for might even need to involve the company board of directors. Following that, the recommendation to practitioners is to not treat continuous experimentation as a project that can be solved with only software development. The whole organization needs to be involved, e.g., to find metrics and ensuring that the user data to measure this can be acquired. Otherwise, if the software development organization conducts experimentation in isolation, the soft aspects of infrastructure might be lacking or the software might be optimized with the wrong goal in mind.

\subsection{Solutions applied (RQ2)}

Concerning the solutions that are applied within continuous experimentation (RQ2), the literature analysis revealed solutions about qualitative feedback, variants of controlled experiments design, quasi-experiments, automated controlled experimentation with optimization algorithms, statistical methods, continuous monitoring, data mining, variability management, and metric specification. For each of these solutions in literature, themes were proposed. One observation made was that the validation of most proposed solutions could be further improved by providing the used data sets, a context description or the necessary steps that allow to reproduce the presented results. Also, many interesting solutions would benefit from further applications that demonstrate their applicability in practice. 
Another observation was that many solutions are driven by practical problems of the author's associated organization (e.g. evaluation of mobile apps \citer{hill2015measuring}). This has the advantage that the problems are of relevance for practice and the provided solutions are assumed to be applicable in similar contexts. Publications of this kind are guidelines for practitioners and valuable research contributions.

There are a lot of solutions for practitioners to choose from, most of them solve a very specific problem that has been observed at a company. In Figure~\ref{fig:rq2-overview}, the solutions are arranged by phase of the experimentation process. What follows is additional help to practitioners to know what solution to apply for a given problem encountered, which is in the design science tradition known as technological rules~\cite{engstrom2020software}:
\begin{itemize}
    \item to achieve \emph{additional insights} in concluded experiments apply \emph{1) data mining} that automatically segments results for users' context;
    \item to achieve \emph{more relevant results} in difficult to measure software systems apply \emph{2) metric specification techniques}.
    \item to achieve \emph{richer experiment feedback} in continuous experimentation apply \emph{3) variants of controlled experiments design} or \emph{9) qualitative feedback}.
    \item to achieve \emph{quantitative results} in environments where parallel deployment is challenging apply \emph{4) quasi-experiments};
    \item to achieve \emph{optimized user interfaces} in software systems that can be evaluated on a single metric apply \emph{5) automated controlled experimentation with optimization algorithms};
    \item to achieve \emph{higher throughput of experiments} in experimentation platforms apply \emph{6) variability management techniques} to specify overlapping experiments;
    \item to achieve \emph{trustworthy results} in online controlled experiments apply \emph{7) improved statistical methods} or \emph{1) data mining} to calibrate the statistical tests;
    \item to achieve \emph{faster results} in online controlled experiments apply \emph{8) continuous monitoring} to help decide when experiments can be stopped early.
\end{itemize}

\subsection{Challenges (RQ3)}

Many authors of the studied literature mentioned challenges with continuous experimentation in their papers.  The thematic analysis of the challenges identified six fundamental challenge themes. Here they are presented along with the recommendations to mitigate the risks.

The \emph{cultural, organizational and managerial} challenges seem to indicate that the multi-disciplinary characteristic of continuous experimentation introduces new requirements to the team. It requires amongst others the collaboration of cross-functional stakeholders (i.e. business, design, and engineering). This can represent a fundamental cultural change within an organization. Hence, the adaption of continuous experimentation involves technical as well as cultural changes. Challenges like the lack of adaption support this interpretation. Mitigating these challenges involves taking a whole organizational approach to continuous experimentation so that both engineers and managers are in agreement about conducting experimentation.

Another theme among challenges is \emph{business}. The challenges assigned to this theme highlight that continuous experimentation has challenges in its economic application with respect to the financial return on investment. The focus of experimentation needs to be managed appropriately in order to prevent investing in incremental development with insufficient impact. Also, that changes cannot be measured with a relevant metric is another business challenge.  
One possible approach for further research on these challenges could be the transfer from solutions in other disciplines to continuous experimentation. An example therefore is the overall evaluation criteria \cite{van2002design} that was adapted to continuous experimentation by Kohavi et al.~\citer{kohavi2007practical}. As with the previous challenge theme, this theme of challenges does not have an easy fix. It might be the case that experimentation is simply not applicable for all software companies but further research is needed to determine this.

Concerning the \emph{technical} challenges, the literature review showed that there are challenges related to continuous deployment/delivery and experiment control. The delivery of changes to production is challenging especially for environments that are used to none or infrequent updates, like embedded devices. For such edge cases, new deployment strategies have to be found that are suitable for continuous experimentation. Although solutions from continuous deployment seem to be fitting, they need to be extended with mechanisms to control the experiment at run-time (e.g. to stop an experiment). This can be challenging in environments for which frequent updates are difficult. There is proof-of-concept research~\citer{eklund2012architecture} to handle these challenges so they do not seem to be impossible blockers to get started on experimentation.

The \emph{statistical} challenges mentioned in the studied literature indicate that there is a need for solutions to cope with the various ways that the statistical assumptions done in a controlled experiment are broken by changes in the real world. There are both changes in the environment (exogenous) and changes caused by experimentation (endogenous). Changes in the environment (e.g. the effect of an advertisement campaign run by the marketing department) can alter the initial situation of an experiment and thus may lead to wrong conclusions about the results. Therefore, the knowledge about an experiment's environment and possible influences needs to be systematically advanced and the experiments themselves should be designed to be more robust. Mitigating these challenges involves identifying and applying the correct solution for the specific problem. There is further research opportunity to document and synthesize such problem-solution pairs.

\emph{Ethical} aspects are not investigated by many studies. The experience reports and lessons learned publications do not, for example, mention user consent or user's awareness of participation. Furthermore, ethical considerations about which experiments should be conducted or not were seldom discussed in the papers. There were still two challenges identified in this study, involving data privacy and dark patterns.
However, examples like the Facebook emotional manipulation study, which changed the user's news feed to determine whether it affects the subsequent posts of a user, show the need for ethical considerations in experimentation \cite{flick2016informed}. Although this was an experiment in the context of an academic study in psychology, the case nevertheless shows that there are open challenges on the topic of ethics and continuous experimentation. There is not enough research conducted for a concrete recommendation other than raising awareness of the existence of ethical dilemmas involving experimentation.

Continuous experimentation is applied in various domains that require \emph{domain specific} solutions. The challenges on continuous experimentation range from infrastructure challenges, over measurement challenges, to social challenges. 
Examples are the challenge to deploy changes in cyber-physical systems (infrastructural challenge), to differentiate the effects of one change from another (measurement challenge), and the influence of users on each other across treatment groups (social challenge).
Each challenge is probably only relevant for certain domains, however the developed solutions may be adaptable to other domains. Thus, the research on domain-specific challenges could take optimized solutions for specific domains to solutions for other domains.

\subsection{Benefits (RQ4)}

In many publications about continuous experimentation the benefits of experimentation are mentioned as motivation only; i.e. it increases the quality of the product based on the chosen metrics. 
The two publications on explicit benefits~\citer{bosch2012building,fabijan2017benefits} mention improvements not only on the product in business-related metrics and usability but also on the product portfolio offering and generic benefits for the whole organization (better collaboration, prioritization, etc.). More studies are needed to determine, e.g., if there are more benefits, whether the benefits apply for all companies involved with experimentation, or whether the benefits could be obtained through other means. 
Another benefit is the potential usage of continuous experimentation for software quality assurance. Continuous experimentation could support or even change the way quality assurance is done for software. Software change, for example, could only be deployed if key metrics are not degraded in the related change experiment. Thus, quality degradation could become quantifiable and measurable. Although some papers, like \citer{fabijan2017benefits}, mention the usage of continuous experimentation for software quality assurance. 

\section{Conclusions}
\label{sec:conclusions}

This paper presents a systematic literature review of the current state of controlled experiments in continuous experimentation. Forward snowballing was applied on the selected paper sets of two previous mapping studies in the field. The 128 papers that were finally selected, were qualitatively analyzed using thematic analysis.

The study found two constituents of a continuous experimentation framework (RQ1): an experimentation process and a supportive infrastructure. Based on experience reports that discuss failed experiments in the context of large-scale software development, the recommendation to practitioners is to apply one of the published processes, but also expand it by placing more emphasis on the ideation phase by making prototypes. As for the infrastructure, several studies discuss requirements for controlled experiments to ramp up the scale and speed of experimentation. Our recommendation for infrastructure is to consider the organizational aspects to ensure that, e.g., the necessary channels for communicating results are in place.

Ten themes of solutions (RQ2) were found that were applied in the various phases of controlled experimentation: data mining, metric specification, variants of controlled experiment design, quasi-experiments, automated controlled experimentation, variability management, continuous monitoring, improved statistical methods, and qualitative feedback. We have provided recommendations on what problem each solution theme solves for what context in the discussion.

Finally, the analysis of challenges (RQ3) and benefits (RQ4) of continuous experimentation revealed that only two papers focused explicitly on the benefits of experimentation. In contrast, multiple papers focused on challenges. The analysis identified six themes of challenges: cultural/organizational, business, technical, statistical, ethical, and domain-specific challenges. While the papers on challenges do outnumber the papers on benefits, there is no cause for concern, as the benefits to product quality are also mentioned in many papers as motivation to conduct the research. The challenges to experimentation also come with recommendations in the discussion on how to mitigate them.

As a final remark, we encourage practitioners to investigate the large body of highly industry-relevant research that exists for controlled experimentation in continuous experimentation and for researchers to follow the many remaining gaps in literature revealed within.

\section*{Acknowledgements}
This work was partially supported by the Wallenberg Artificial Intelligence, Autonomous Systems and Software Program (WASP) funded by Knut and Alice Wallenberg Foundation and the Austrian Science Fund (FWF): I 4701-N.

\section*{\refname}
\bibliographystyle{elsarticle-num}
\bibliography{allreferences}

\section*{Selected Publications}
\nociter{*}
\bibliographystyler{elsarticle-num}
\bibliographyr{selectedpapers}


\end{document}

%% file: figure-ce-process.tex
\ifx\du\undefined
  \newlength{\du}
\fi
\setlength{\du}{15\unitlength}
\begin{tikzpicture}[even odd rule]
\pgftransformxscale{1.000000}
\pgftransformyscale{-1.000000}
\definecolor{dialinecolor}{rgb}{0.000000, 0.000000, 0.000000}
\pgfsetstrokecolor{dialinecolor}
\pgfsetstrokeopacity{1.000000}
\definecolor{diafillcolor}{rgb}{1.000000, 1.000000, 1.000000}
\pgfsetfillcolor{diafillcolor}
\pgfsetfillopacity{1.000000}
\pgfsetlinewidth{0.050000\du}
\pgfsetdash{}{0pt}
\pgfsetmiterjoin
{\pgfsetcornersarced{\pgfpoint{0.000000\du}{0.000000\du}}\definecolor{diafillcolor}{rgb}{1.000000, 1.000000, 1.000000}
\pgfsetfillcolor{diafillcolor}
\pgfsetfillopacity{1.000000}
\fill (14.400000\du,10.800000\du)--(14.400000\du,12.343889\du)--(21.900000\du,12.343889\du)--(21.900000\du,10.800000\du)--cycle;
}{\pgfsetcornersarced{\pgfpoint{0.000000\du}{0.000000\du}}\definecolor{dialinecolor}{rgb}{0.000000, 0.000000, 0.000000}
\pgfsetstrokecolor{dialinecolor}
\pgfsetstrokeopacity{1.000000}
\draw (14.400000\du,10.800000\du)--(14.400000\du,12.343889\du)--(21.900000\du,12.343889\du)--(21.900000\du,10.800000\du)--cycle;
}
\definecolor{dialinecolor}{rgb}{0.000000, 0.000000, 0.000000}
\pgfsetstrokecolor{dialinecolor}
\pgfsetstrokeopacity{1.000000}
\definecolor{diafillcolor}{rgb}{0.000000, 0.000000, 0.000000}
\pgfsetfillcolor{diafillcolor}
\pgfsetfillopacity{1.000000}
\node[anchor=base,inner sep=0pt, outer sep=0pt,color=dialinecolor] at (18.150000\du,11.692500\du){1) Ideation};
\pgfsetlinewidth{0.050000\du}
\pgfsetdash{}{0pt}
\pgfsetmiterjoin
{\pgfsetcornersarced{\pgfpoint{0.000000\du}{0.000000\du}}\definecolor{diafillcolor}{rgb}{1.000000, 1.000000, 1.000000}
\pgfsetfillcolor{diafillcolor}
\pgfsetfillopacity{1.000000}
\fill (14.400000\du,13.400000\du)--(14.400000\du,14.943889\du)--(21.900000\du,14.943889\du)--(21.900000\du,13.400000\du)--cycle;
}{\pgfsetcornersarced{\pgfpoint{0.000000\du}{0.000000\du}}\definecolor{dialinecolor}{rgb}{0.000000, 0.000000, 0.000000}
\pgfsetstrokecolor{dialinecolor}
\pgfsetstrokeopacity{1.000000}
\draw (14.400000\du,13.400000\du)--(14.400000\du,14.943889\du)--(21.900000\du,14.943889\du)--(21.900000\du,13.400000\du)--cycle;
}
\definecolor{dialinecolor}{rgb}{0.000000, 0.000000, 0.000000}
\pgfsetstrokecolor{dialinecolor}
\pgfsetstrokeopacity{1.000000}
\definecolor{diafillcolor}{rgb}{0.000000, 0.000000, 0.000000}
\pgfsetfillcolor{diafillcolor}
\pgfsetfillopacity{1.000000}
\node[anchor=base,inner sep=0pt, outer sep=0pt,color=dialinecolor] at (18.150000\du,14.292500\du){2) Implementation};
\pgfsetlinewidth{0.050000\du}
\pgfsetdash{}{0pt}
\pgfsetmiterjoin
{\pgfsetcornersarced{\pgfpoint{0.000000\du}{0.000000\du}}\definecolor{diafillcolor}{rgb}{1.000000, 1.000000, 1.000000}
\pgfsetfillcolor{diafillcolor}
\pgfsetfillopacity{1.000000}
\fill (14.400000\du,16.000000\du)--(14.400000\du,17.543889\du)--(21.900000\du,17.543889\du)--(21.900000\du,16.000000\du)--cycle;
}{\pgfsetcornersarced{\pgfpoint{0.000000\du}{0.000000\du}}\definecolor{dialinecolor}{rgb}{0.000000, 0.000000, 0.000000}
\pgfsetstrokecolor{dialinecolor}
\pgfsetstrokeopacity{1.000000}
\draw (14.400000\du,16.000000\du)--(14.400000\du,17.543889\du)--(21.900000\du,17.543889\du)--(21.900000\du,16.000000\du)--cycle;
}
\definecolor{dialinecolor}{rgb}{0.000000, 0.000000, 0.000000}
\pgfsetstrokecolor{dialinecolor}
\pgfsetstrokeopacity{1.000000}
\definecolor{diafillcolor}{rgb}{0.000000, 0.000000, 0.000000}
\pgfsetfillcolor{diafillcolor}
\pgfsetfillopacity{1.000000}
\node[anchor=base,inner sep=0pt, outer sep=0pt,color=dialinecolor] at (18.150000\du,16.892500\du){3) Experiment design};
\pgfsetlinewidth{0.050000\du}
\pgfsetdash{}{0pt}
\pgfsetmiterjoin
{\pgfsetcornersarced{\pgfpoint{0.000000\du}{0.000000\du}}\definecolor{diafillcolor}{rgb}{1.000000, 1.000000, 1.000000}
\pgfsetfillcolor{diafillcolor}
\pgfsetfillopacity{1.000000}
\fill (14.400000\du,18.600000\du)--(14.400000\du,20.143889\du)--(21.900000\du,20.143889\du)--(21.900000\du,18.600000\du)--cycle;
}{\pgfsetcornersarced{\pgfpoint{0.000000\du}{0.000000\du}}\definecolor{dialinecolor}{rgb}{0.000000, 0.000000, 0.000000}
\pgfsetstrokecolor{dialinecolor}
\pgfsetstrokeopacity{1.000000}
\draw (14.400000\du,18.600000\du)--(14.400000\du,20.143889\du)--(21.900000\du,20.143889\du)--(21.900000\du,18.600000\du)--cycle;
}
\definecolor{dialinecolor}{rgb}{0.000000, 0.000000, 0.000000}
\pgfsetstrokecolor{dialinecolor}
\pgfsetstrokeopacity{1.000000}
\definecolor{diafillcolor}{rgb}{0.000000, 0.000000, 0.000000}
\pgfsetfillcolor{diafillcolor}
\pgfsetfillopacity{1.000000}
\node[anchor=base,inner sep=0pt, outer sep=0pt,color=dialinecolor] at (18.150000\du,19.492500\du){4) Execution};
\pgfsetlinewidth{0.050000\du}
\pgfsetdash{}{0pt}
\pgfsetbuttcap
{
\definecolor{diafillcolor}{rgb}{0.000000, 0.000000, 0.000000}
\pgfsetfillcolor{diafillcolor}
\pgfsetfillopacity{1.000000}
\pgfsetarrowsend{stealth}
\definecolor{dialinecolor}{rgb}{0.000000, 0.000000, 0.000000}
\pgfsetstrokecolor{dialinecolor}
\pgfsetstrokeopacity{1.000000}
\draw (18.150000\du,12.366671\du)--(18.150000\du,13.377218\du);
}
\pgfsetlinewidth{0.050000\du}
\pgfsetdash{}{0pt}
\pgfsetbuttcap
{
\definecolor{diafillcolor}{rgb}{0.000000, 0.000000, 0.000000}
\pgfsetfillcolor{diafillcolor}
\pgfsetfillopacity{1.000000}
\pgfsetarrowsend{stealth}
\definecolor{dialinecolor}{rgb}{0.000000, 0.000000, 0.000000}
\pgfsetstrokecolor{dialinecolor}
\pgfsetstrokeopacity{1.000000}
\draw (18.150000\du,14.943889\du)--(18.150000\du,16.000000\du);
}
\pgfsetlinewidth{0.050000\du}
\pgfsetdash{}{0pt}
\pgfsetbuttcap
{
\definecolor{diafillcolor}{rgb}{0.000000, 0.000000, 0.000000}
\pgfsetfillcolor{diafillcolor}
\pgfsetfillopacity{1.000000}
\pgfsetarrowsend{stealth}
\definecolor{dialinecolor}{rgb}{0.000000, 0.000000, 0.000000}
\pgfsetstrokecolor{dialinecolor}
\pgfsetstrokeopacity{1.000000}
\draw (18.150000\du,17.543889\du)--(18.150000\du,18.600000\du);
}
\pgfsetlinewidth{0.050000\du}
\pgfsetdash{}{0pt}
\pgfsetbuttcap
{
\definecolor{diafillcolor}{rgb}{0.000000, 0.000000, 0.000000}
\pgfsetfillcolor{diafillcolor}
\pgfsetfillopacity{1.000000}
\pgfsetarrowsend{stealth}
\definecolor{dialinecolor}{rgb}{0.000000, 0.000000, 0.000000}
\pgfsetstrokecolor{dialinecolor}
\pgfsetstrokeopacity{1.000000}
\draw (21.900000\du,19.371944\du)--(23.600000\du,19.400000\du);
}
\pgfsetlinewidth{0.050000\du}
\pgfsetdash{}{0pt}
\pgfsetmiterjoin
\definecolor{diafillcolor}{rgb}{1.000000, 1.000000, 1.000000}
\pgfsetfillcolor{diafillcolor}
\pgfsetfillopacity{1.000000}
\fill (26.200000\du,18.000000\du)--(28.800000\du,19.400000\du)--(26.200000\du,20.800000\du)--(23.600000\du,19.400000\du)--cycle;
\definecolor{dialinecolor}{rgb}{0.000000, 0.000000, 0.000000}
\pgfsetstrokecolor{dialinecolor}
\pgfsetstrokeopacity{1.000000}
\draw (26.200000\du,18.000000\du)--(28.800000\du,19.400000\du)--(26.200000\du,20.800000\du)--(23.600000\du,19.400000\du)--cycle;
\definecolor{dialinecolor}{rgb}{0.000000, 0.000000, 0.000000}
\pgfsetstrokecolor{dialinecolor}
\pgfsetstrokeopacity{1.000000}
\definecolor{diafillcolor}{rgb}{0.000000, 0.000000, 0.000000}
\pgfsetfillcolor{diafillcolor}
\pgfsetfillopacity{1.000000}
\node[anchor=base,inner sep=0pt, outer sep=0pt,color=dialinecolor] at (26.200000\du,19.595000\du){5) Analysis};
\definecolor{dialinecolor}{rgb}{0.000000, 0.000000, 0.000000}
\pgfsetstrokecolor{dialinecolor}
\pgfsetstrokeopacity{1.000000}
\definecolor{diafillcolor}{rgb}{0.000000, 0.000000, 0.000000}
\pgfsetfillcolor{diafillcolor}
\pgfsetfillopacity{1.000000}
\node[anchor=base west,inner sep=0pt,outer sep=0pt,color=dialinecolor] at (29.000000\du,19.000000\du){Pivot};
\pgfsetlinewidth{0.050000\du}
\pgfsetdash{}{0pt}
\pgfsetmiterjoin
\pgfsetbuttcap
{
\definecolor{diafillcolor}{rgb}{0.000000, 0.000000, 0.000000}
\pgfsetfillcolor{diafillcolor}
\pgfsetfillopacity{1.000000}
\pgfsetarrowsend{stealth}
{\pgfsetcornersarced{\pgfpoint{0.000000\du}{0.000000\du}}\definecolor{dialinecolor}{rgb}{0.000000, 0.000000, 0.000000}
\pgfsetstrokecolor{dialinecolor}
\pgfsetstrokeopacity{1.000000}
\draw (26.200000\du,18.000000\du)--(26.200000\du,18.000000\du)--(26.200000\du,11.957917\du)--(21.900000\du,11.957917\du);
}}
\definecolor{dialinecolor}{rgb}{0.000000, 0.000000, 0.000000}
\pgfsetstrokecolor{dialinecolor}
\pgfsetstrokeopacity{1.000000}
\definecolor{diafillcolor}{rgb}{0.000000, 0.000000, 0.000000}
\pgfsetfillcolor{diafillcolor}
\pgfsetfillopacity{1.000000}
\node[anchor=base west,inner sep=0pt,outer sep=0pt,color=dialinecolor] at (22.200000\du,11.600000\du){Refine / perservere};
\pgfsetlinewidth{0.050000\du}
\pgfsetdash{}{0pt}
\pgfsetbuttcap
{
\definecolor{diafillcolor}{rgb}{0.000000, 0.000000, 0.000000}
\pgfsetfillcolor{diafillcolor}
\pgfsetfillopacity{1.000000}
\pgfsetarrowsend{stealth}
\definecolor{dialinecolor}{rgb}{0.000000, 0.000000, 0.000000}
\pgfsetstrokecolor{dialinecolor}
\pgfsetstrokeopacity{1.000000}
\draw (28.800000\du,19.400000\du)--(30.600000\du,19.400000\du);
}
\definecolor{dialinecolor}{rgb}{0.000000, 0.000000, 0.000000}
\pgfsetstrokecolor{dialinecolor}
\pgfsetstrokeopacity{1.000000}
\definecolor{diafillcolor}{rgb}{0.000000, 0.000000, 0.000000}
\pgfsetfillcolor{diafillcolor}
\pgfsetfillopacity{1.000000}
\node[anchor=base west,inner sep=0pt,outer sep=0pt,color=dialinecolor] at (8.800000\du,11.400000\du){Business analyst};
\definecolor{dialinecolor}{rgb}{0.000000, 0.000000, 0.000000}
\pgfsetstrokecolor{dialinecolor}
\pgfsetstrokeopacity{1.000000}
\definecolor{diafillcolor}{rgb}{0.000000, 0.000000, 0.000000}
\pgfsetfillcolor{diafillcolor}
\pgfsetfillopacity{1.000000}
\node[anchor=base west,inner sep=0pt,outer sep=0pt,color=dialinecolor] at (8.800000\du,12.200000\du){Product owner};
\definecolor{dialinecolor}{rgb}{0.000000, 0.000000, 0.000000}
\pgfsetstrokecolor{dialinecolor}
\pgfsetstrokeopacity{1.000000}
\definecolor{diafillcolor}{rgb}{0.000000, 0.000000, 0.000000}
\pgfsetfillcolor{diafillcolor}
\pgfsetfillopacity{1.000000}
\node[anchor=base west,inner sep=0pt,outer sep=0pt,color=dialinecolor] at (8.800000\du,14.000000\du){Software developer};
\definecolor{dialinecolor}{rgb}{0.000000, 0.000000, 0.000000}
\pgfsetstrokecolor{dialinecolor}
\pgfsetstrokeopacity{1.000000}
\definecolor{diafillcolor}{rgb}{0.000000, 0.000000, 0.000000}
\pgfsetfillcolor{diafillcolor}
\pgfsetfillopacity{1.000000}
\node[anchor=base west,inner sep=0pt,outer sep=0pt,color=dialinecolor] at (8.800000\du,14.800000\du){Quality assurance};
\definecolor{dialinecolor}{rgb}{0.000000, 0.000000, 0.000000}
\pgfsetstrokecolor{dialinecolor}
\pgfsetstrokeopacity{1.000000}
\definecolor{diafillcolor}{rgb}{0.000000, 0.000000, 0.000000}
\pgfsetfillcolor{diafillcolor}
\pgfsetfillopacity{1.000000}
\node[anchor=base west,inner sep=0pt,outer sep=0pt,color=dialinecolor] at (8.800000\du,16.600000\du){Data scientist};
\definecolor{dialinecolor}{rgb}{0.000000, 0.000000, 0.000000}
\pgfsetstrokecolor{dialinecolor}
\pgfsetstrokeopacity{1.000000}
\definecolor{diafillcolor}{rgb}{0.000000, 0.000000, 0.000000}
\pgfsetfillcolor{diafillcolor}
\pgfsetfillopacity{1.000000}
\node[anchor=base west,inner sep=0pt,outer sep=0pt,color=dialinecolor] at (8.800000\du,17.400000\du){User researcher};
\definecolor{dialinecolor}{rgb}{0.000000, 0.000000, 0.000000}
\pgfsetstrokecolor{dialinecolor}
\pgfsetstrokeopacity{1.000000}
\definecolor{diafillcolor}{rgb}{0.000000, 0.000000, 0.000000}
\pgfsetfillcolor{diafillcolor}
\pgfsetfillopacity{1.000000}
\node[anchor=base west,inner sep=0pt,outer sep=0pt,color=dialinecolor] at (8.800000\du,19.200000\du){Release engineer};
\definecolor{dialinecolor}{rgb}{0.000000, 0.000000, 0.000000}
\pgfsetstrokecolor{dialinecolor}
\pgfsetstrokeopacity{1.000000}
\definecolor{diafillcolor}{rgb}{0.000000, 0.000000, 0.000000}
\pgfsetfillcolor{diafillcolor}
\pgfsetfillopacity{1.000000}
\node[anchor=base west,inner sep=0pt,outer sep=0pt,color=dialinecolor] at (8.800000\du,20.000000\du){Operations engineer};
\end{tikzpicture}

%% file: solutions.tex
\ifx\du\undefined
  \newlength{\du}
\fi
\setlength{\du}{15\unitlength}
\begin{tikzpicture}[even odd rule]
\pgftransformxscale{1.000000}
\pgftransformyscale{-1.000000}
\definecolor{dialinecolor}{rgb}{0.000000, 0.000000, 0.000000}
\pgfsetstrokecolor{dialinecolor}
\pgfsetstrokeopacity{1.000000}
\definecolor{diafillcolor}{rgb}{1.000000, 1.000000, 1.000000}
\pgfsetfillcolor{diafillcolor}
\pgfsetfillopacity{1.000000}
\pgfsetlinewidth{0.050000\du}
\pgfsetdash{}{0pt}
\pgfsetmiterjoin
{\pgfsetcornersarced{\pgfpoint{0.400000\du}{0.400000\du}}\definecolor{diafillcolor}{rgb}{1.000000, 1.000000, 1.000000}
\pgfsetfillcolor{diafillcolor}
\pgfsetfillopacity{1.000000}
\fill (6.000000\du,3.000000\du)--(6.000000\du,6.500000\du)--(15.000000\du,6.500000\du)--(15.000000\du,3.000000\du)--cycle;
}{\pgfsetcornersarced{\pgfpoint{0.400000\du}{0.400000\du}}\definecolor{dialinecolor}{rgb}{0.000000, 0.000000, 0.000000}
\pgfsetstrokecolor{dialinecolor}
\pgfsetstrokeopacity{1.000000}
\draw (6.000000\du,3.000000\du)--(6.000000\du,6.500000\du)--(15.000000\du,6.500000\du)--(15.000000\du,3.000000\du)--cycle;
}
\definecolor{dialinecolor}{rgb}{0.000000, 0.000000, 0.000000}
\pgfsetstrokecolor{dialinecolor}
\pgfsetstrokeopacity{1.000000}
\definecolor{diafillcolor}{rgb}{0.000000, 0.000000, 0.000000}
\pgfsetfillcolor{diafillcolor}
\pgfsetfillopacity{1.000000}
\node[anchor=base west,inner sep=0pt,outer sep=0pt,color=dialinecolor] at (6.475000\du,3.952500\du){\normalsize{Analysis}};
\definecolor{dialinecolor}{rgb}{0.000000, 0.000000, 0.000000}
\pgfsetstrokecolor{dialinecolor}
\pgfsetstrokeopacity{1.000000}
\definecolor{diafillcolor}{rgb}{0.000000, 0.000000, 0.000000}
\pgfsetfillcolor{diafillcolor}
\pgfsetfillopacity{1.000000}
\node[anchor=base west,inner sep=0pt,outer sep=0pt,color=dialinecolor] at (6.475000\du,4.787500\du){8. Improved statistical};
\definecolor{dialinecolor}{rgb}{0.000000, 0.000000, 0.000000}
\pgfsetstrokecolor{dialinecolor}
\pgfsetstrokeopacity{1.000000}
\definecolor{diafillcolor}{rgb}{0.000000, 0.000000, 0.000000}
\pgfsetfillcolor{diafillcolor}
\pgfsetfillopacity{1.000000}
\node[anchor=base west,inner sep=0pt,outer sep=0pt,color=dialinecolor] at (7.225\du,5.422500\du){  methods};
\definecolor{dialinecolor}{rgb}{0.000000, 0.000000, 0.000000}
\pgfsetstrokecolor{dialinecolor}
\pgfsetstrokeopacity{1.000000}
\definecolor{diafillcolor}{rgb}{0.000000, 0.000000, 0.000000}
\pgfsetfillcolor{diafillcolor}
\pgfsetfillopacity{1.000000}
\node[anchor=base west,inner sep=0pt,outer sep=0pt,color=dialinecolor] at (6.475000\du,6.057500\du){9. Qualitative feedback};
\pgfsetlinewidth{0.050000\du}
\pgfsetdash{}{0pt}
\pgfsetmiterjoin
{\pgfsetcornersarced{\pgfpoint{0.400000\du}{0.400000\du}}\definecolor{diafillcolor}{rgb}{1.000000, 1.000000, 1.000000}
\pgfsetfillcolor{diafillcolor}
\pgfsetfillopacity{1.000000}
\fill (12.000000\du,-1.500000\du)--(12.000000\du,2.000000\du)--(21.000000\du,2.000000\du)--(21.000000\du,-1.500000\du)--cycle;
}{\pgfsetcornersarced{\pgfpoint{0.400000\du}{0.400000\du}}\definecolor{dialinecolor}{rgb}{0.000000, 0.000000, 0.000000}
\pgfsetstrokecolor{dialinecolor}
\pgfsetstrokeopacity{1.000000}
\draw (12.000000\du,-1.500000\du)--(12.000000\du,2.000000\du)--(21.000000\du,2.000000\du)--(21.000000\du,-1.500000\du)--cycle;
}
\definecolor{dialinecolor}{rgb}{0.000000, 0.000000, 0.000000}
\pgfsetstrokecolor{dialinecolor}
\pgfsetstrokeopacity{1.000000}
\definecolor{diafillcolor}{rgb}{0.000000, 0.000000, 0.000000}
\pgfsetfillcolor{diafillcolor}
\pgfsetfillopacity{1.000000}
\node[anchor=base west,inner sep=0pt,outer sep=0pt,color=dialinecolor] at (12.475000\du,-0.50000\du){\normalsize{Ideation}};
\definecolor{dialinecolor}{rgb}{0.000000, 0.000000, 0.000000}
\pgfsetstrokecolor{dialinecolor}
\pgfsetstrokeopacity{1.000000}
\definecolor{diafillcolor}{rgb}{0.000000, 0.000000, 0.000000}
\pgfsetfillcolor{diafillcolor}
\pgfsetfillopacity{1.000000}
\node[anchor=base west,inner sep=0pt,outer sep=0pt,color=dialinecolor] at (12.475000\du,0.425000\du){1. Data mining};
\definecolor{dialinecolor}{rgb}{0.000000, 0.000000, 0.000000}
\pgfsetstrokecolor{dialinecolor}
\pgfsetstrokeopacity{1.000000}
\definecolor{diafillcolor}{rgb}{0.000000, 0.000000, 0.000000}
\pgfsetfillcolor{diafillcolor}
\pgfsetfillopacity{1.000000}
\node[anchor=base west,inner sep=0pt,outer sep=0pt,color=dialinecolor] at (12.475000\du,1.240000\du){2. Metric specification};
\pgfsetlinewidth{0.050000\du}
\pgfsetdash{}{0pt}
\pgfsetmiterjoin
{\pgfsetcornersarced{\pgfpoint{0.400000\du}{0.400000\du}}\definecolor{diafillcolor}{rgb}{1.000000, 1.000000, 1.000000}
\pgfsetfillcolor{diafillcolor}
\pgfsetfillopacity{1.000000}
\fill (18.000000\du,3.000000\du)--(18.000000\du,6.500000\du)--(27.000000\du,6.500000\du)--(27.000000\du,3.000000\du)--cycle;
}{\pgfsetcornersarced{\pgfpoint{0.400000\du}{0.400000\du}}\definecolor{dialinecolor}{rgb}{0.000000, 0.000000, 0.000000}
\pgfsetstrokecolor{dialinecolor}
\pgfsetstrokeopacity{1.000000}
\draw (18.000000\du,3.000000\du)--(18.000000\du,6.500000\du)--(27.000000\du,6.500000\du)--(27.000000\du,3.000000\du)--cycle;
}
\definecolor{dialinecolor}{rgb}{0.000000, 0.000000, 0.000000}
\pgfsetstrokecolor{dialinecolor}
\pgfsetstrokeopacity{1.000000}
\definecolor{diafillcolor}{rgb}{0.000000, 0.000000, 0.000000}
\pgfsetfillcolor{diafillcolor}
\pgfsetfillopacity{1.000000}
\node[anchor=base west,inner sep=0pt,outer sep=0pt,color=dialinecolor] at (18.475000\du,3.952500\du){\normalsize{Experiment design}};
\definecolor{dialinecolor}{rgb}{0.000000, 0.000000, 0.000000}
\pgfsetstrokecolor{dialinecolor}
\pgfsetstrokeopacity{1.000000}
\definecolor{diafillcolor}{rgb}{0.000000, 0.000000, 0.000000}
\pgfsetfillcolor{diafillcolor}
\pgfsetfillopacity{1.000000}
\node[anchor=base west,inner sep=0pt,outer sep=0pt,color=dialinecolor] at (18.475000\du,4.787500\du){3. Variants of controlled };
\definecolor{dialinecolor}{rgb}{0.000000, 0.000000, 0.000000}
\pgfsetstrokecolor{dialinecolor}
\pgfsetstrokeopacity{1.000000}
\definecolor{diafillcolor}{rgb}{0.000000, 0.000000, 0.000000}
\pgfsetfillcolor{diafillcolor}
\pgfsetfillopacity{1.000000}
\node[anchor=base west,inner sep=0pt,outer sep=0pt,color=dialinecolor] at (19.270000\du,5.422500\du){   experiment design};
\definecolor{dialinecolor}{rgb}{0.000000, 0.000000, 0.000000}
\pgfsetstrokecolor{dialinecolor}
\pgfsetstrokeopacity{1.000000}
\definecolor{diafillcolor}{rgb}{0.000000, 0.000000, 0.000000}
\pgfsetfillcolor{diafillcolor}
\pgfsetfillopacity{1.000000}
\node[anchor=base west,inner sep=0pt,outer sep=0pt,color=dialinecolor] at (18.475000\du,6.057500\du){4. Quasi experiments};
\pgfsetlinewidth{0.050000\du}
\pgfsetdash{}{0pt}
\pgfsetmiterjoin
{\pgfsetcornersarced{\pgfpoint{0.400000\du}{0.400000\du}}\definecolor{diafillcolor}{rgb}{1.000000, 1.000000, 1.000000}
\pgfsetfillcolor{diafillcolor}
\pgfsetfillopacity{1.000000}
\fill (6.000000\du,8.000000\du)--(6.000000\du,11.500000\du)--(15.000000\du,11.500000\du)--(15.000000\du,8.000000\du)--cycle;
}{\pgfsetcornersarced{\pgfpoint{0.400000\du}{0.400000\du}}\definecolor{dialinecolor}{rgb}{0.000000, 0.000000, 0.000000}
\pgfsetstrokecolor{dialinecolor}
\pgfsetstrokeopacity{1.000000}
\draw (6.000000\du,8.000000\du)--(6.000000\du,11.500000\du)--(15.000000\du,11.500000\du)--(15.000000\du,8.000000\du)--cycle;
}
\definecolor{dialinecolor}{rgb}{0.000000, 0.000000, 0.000000}
\pgfsetstrokecolor{dialinecolor}
\pgfsetstrokeopacity{1.000000}
\definecolor{diafillcolor}{rgb}{0.000000, 0.000000, 0.000000}
\pgfsetfillcolor{diafillcolor}
\pgfsetfillopacity{1.000000}
\node[anchor=base west,inner sep=0pt,outer sep=0pt,color=dialinecolor] at (6.475000\du,8.9525\du){\normalsize{Execution}};
\definecolor{dialinecolor}{rgb}{0.000000, 0.000000, 0.000000}
\pgfsetstrokecolor{dialinecolor}
\pgfsetstrokeopacity{1.000000}
\definecolor{diafillcolor}{rgb}{0.000000, 0.000000, 0.000000}
\pgfsetfillcolor{diafillcolor}
\pgfsetfillopacity{1.000000}
\node[anchor=base west,inner sep=0pt,outer sep=0pt,color=dialinecolor] at (6.475000\du,10.222500\du){7. Continuous monitoring};
\pgfsetlinewidth{0.050000\du}
\pgfsetdash{}{0pt}
\pgfsetmiterjoin
{\pgfsetcornersarced{\pgfpoint{0.400000\du}{0.400000\du}}\definecolor{diafillcolor}{rgb}{1.000000, 1.000000, 1.000000}
\pgfsetfillcolor{diafillcolor}
\pgfsetfillopacity{1.000000}
\fill (18.000000\du,8.000000\du)--(18.000000\du,11.500000\du)--(27.000000\du,11.500000\du)--(27.000000\du,8.000000\du)--cycle;
}{\pgfsetcornersarced{\pgfpoint{0.400000\du}{0.400000\du}}\definecolor{dialinecolor}{rgb}{0.000000, 0.000000, 0.000000}
\pgfsetstrokecolor{dialinecolor}
\pgfsetstrokeopacity{1.000000}
\draw (18.000000\du,8.000000\du)--(18.000000\du,11.500000\du)--(27.000000\du,11.500000\du)--(27.000000\du,8.000000\du)--cycle;
}
\definecolor{dialinecolor}{rgb}{0.000000, 0.000000, 0.000000}
\pgfsetstrokecolor{dialinecolor}
\pgfsetstrokeopacity{1.000000}
\definecolor{diafillcolor}{rgb}{0.000000, 0.000000, 0.000000}
\pgfsetfillcolor{diafillcolor}
\pgfsetfillopacity{1.000000}
\node[anchor=base west,inner sep=0pt,outer sep=0pt,color=dialinecolor] at (18.475000\du,8.952500\du){\normalsize{Implementation}};
\definecolor{dialinecolor}{rgb}{0.000000, 0.000000, 0.000000}
\pgfsetstrokecolor{dialinecolor}
\pgfsetstrokeopacity{1.000000}
\definecolor{diafillcolor}{rgb}{0.000000, 0.000000, 0.000000}
\pgfsetfillcolor{diafillcolor}
\pgfsetfillopacity{1.000000}
\node[anchor=base west,inner sep=0pt,outer sep=0pt,color=dialinecolor] at (18.475000\du,9.787500\du){5. Automated controlled };
\definecolor{dialinecolor}{rgb}{0.000000, 0.000000, 0.000000}
\pgfsetstrokecolor{dialinecolor}
\pgfsetstrokeopacity{1.000000}
\definecolor{diafillcolor}{rgb}{0.000000, 0.000000, 0.000000}
\pgfsetfillcolor{diafillcolor}
\pgfsetfillopacity{1.000000}
\node[anchor=base west,inner sep=0pt,outer sep=0pt,color=dialinecolor] at (19.27000\du,10.422500\du){  experimentation};
\definecolor{dialinecolor}{rgb}{0.000000, 0.000000, 0.000000}
\pgfsetstrokecolor{dialinecolor}
\pgfsetstrokeopacity{1.000000}
\definecolor{diafillcolor}{rgb}{0.000000, 0.000000, 0.000000}
\pgfsetfillcolor{diafillcolor}
\pgfsetfillopacity{1.000000}
\node[anchor=base west,inner sep=0pt,outer sep=0pt,color=dialinecolor] at (18.475000\du,11.057500\du){6. Variability management};
\pgfsetlinewidth{0.080000\du}
\pgfsetdash{}{0pt}
\pgfsetbuttcap
{
\definecolor{diafillcolor}{rgb}{0.000000, 0.000000, 0.000000}
\pgfsetfillcolor{diafillcolor}
\pgfsetfillopacity{1.000000}
\pgfsetarrowsstart{stealth}
\definecolor{dialinecolor}{rgb}{0.000000, 0.000000, 0.000000}
\pgfsetstrokecolor{dialinecolor}
\pgfsetstrokeopacity{1.000000}
\pgfpathmoveto{\pgfpoint{11.000100\du}{-0.000036\du}}
\pgfpatharc{251}{200}{3.308712\du and 3.308712\du}
\pgfusepath{stroke}
}
\pgfsetlinewidth{0.080000\du}
\pgfsetdash{}{0pt}
\pgfsetbuttcap
{
\definecolor{diafillcolor}{rgb}{0.000000, 0.000000, 0.000000}
\pgfsetfillcolor{diafillcolor}
\pgfsetfillopacity{1.000000}
\pgfsetarrowsstart{stealth}
\definecolor{dialinecolor}{rgb}{0.000000, 0.000000, 0.000000}
\pgfsetstrokecolor{dialinecolor}
\pgfsetstrokeopacity{1.000000}
\pgfpathmoveto{\pgfpoint{24.000006\du}{2.000018\du}}
\pgfpatharc{341}{290}{3.308712\du and 3.308712\du}
\pgfusepath{stroke}
}
\pgfsetlinewidth{0.080000\du}
\pgfsetdash{}{0pt}
\pgfsetbuttcap
{
\definecolor{diafillcolor}{rgb}{0.000000, 0.000000, 0.000000}
\pgfsetfillcolor{diafillcolor}
\pgfsetfillopacity{1.000000}
\pgfsetarrowsstart{stealth}
\definecolor{dialinecolor}{rgb}{0.000000, 0.000000, 0.000000}
\pgfsetstrokecolor{dialinecolor}
\pgfsetstrokeopacity{1.000000}
\pgfpathmoveto{\pgfpoint{23.599864\du}{7.800051\du}}
\pgfpatharc{70}{21}{1.707107\du and 1.707107\du}
\pgfusepath{stroke}
}
\pgfsetlinewidth{0.080000\du}
\pgfsetdash{}{0pt}
\pgfsetbuttcap
{
\definecolor{diafillcolor}{rgb}{0.000000, 0.000000, 0.000000}
\pgfsetfillcolor{diafillcolor}
\pgfsetfillopacity{1.000000}
\pgfsetarrowsstart{stealth}
\definecolor{dialinecolor}{rgb}{0.000000, 0.000000, 0.000000}
\pgfsetstrokecolor{dialinecolor}
\pgfsetstrokeopacity{1.000000}
\pgfpathmoveto{\pgfpoint{15.499847\du}{10.499928\du}}
\pgfpatharc{116}{65}{2.353671\du and 2.353671\du}
\pgfusepath{stroke}
}
\pgfsetlinewidth{0.080000\du}
\pgfsetdash{}{0pt}
\pgfsetbuttcap
{
\definecolor{diafillcolor}{rgb}{0.000000, 0.000000, 0.000000}
\pgfsetfillcolor{diafillcolor}
\pgfsetfillopacity{1.000000}
\pgfsetarrowsstart{stealth}
\definecolor{dialinecolor}{rgb}{0.000000, 0.000000, 0.000000}
\pgfsetstrokecolor{dialinecolor}
\pgfsetstrokeopacity{1.000000}
\pgfpathmoveto{\pgfpoint{8.399965\du}{6.799906\du}}
\pgfpatharc{160}{111}{1.707107\du and 1.707107\du}
\pgfusepath{stroke}
}
\end{tikzpicture}